\documentclass[aps,twocolumn,showpacs,superscriptaddress]{revtex4-1} 
\usepackage[english]{babel}
\usepackage{amsmath,amssymb,amsfonts,bm,graphicx,verbatim,mathrsfs,units,color}
\usepackage[colorlinks=true,citecolor=blue]{hyperref}

\newcommand{\secs}[1]{{\textit{#1.---}}}

\begin{document}
	
\title{Universal First-Passage-Time Distribution of Non-Gaussian Currents}

\author{Shilpi Singh}
 \affiliation{QTF Centre of Excellence, Department of Applied Physics, Aalto University, 00076 Aalto, Finland}

\author{Paul Menczel}
 \affiliation{QTF Centre of Excellence, Department of Applied Physics, Aalto University, 00076 Aalto, Finland}

\author{Dmitry S. Golubev}
 \affiliation{QTF Centre of Excellence, Department of Applied Physics, Aalto University, 00076 Aalto, Finland}

\author{Ivan M. Khaymovich}
  \affiliation{Max Planck Institute for the Physics of Complex Systems, N\"othnitzer Strasse 38, 01187 Dresden, Germany}
  \affiliation{Institute for Physics of Microstructures, Russian Academy of Sciences, 603950 Nizhny Novgorod, GSP-105, Russia}
  
\author{Joonas T. Peltonen}
  \affiliation{QTF Centre of Excellence, Department of Applied Physics, Aalto University, 00076 Aalto, Finland}

\author{Christian Flindt}
 \affiliation{QTF Centre of Excellence, Department of Applied Physics, Aalto University, 00076 Aalto, Finland}

\author{Keiji Saito}
 \affiliation{Department of Physics, Keio University - Yokohama 223–8522, Japan}

\author{ \'Edgar Rold\'an}
   \affiliation{The Abdus Salam International Centre for Theoretical Physics, Strada Costiera 11, 34151, Trieste, Italy}

\author{Jukka P. Pekola}
 \affiliation{QTF Centre of Excellence, Department of Applied Physics, Aalto University, 00076 Aalto, Finland}

\date{\today}

\begin{abstract}

We investigate the fluctuations of the time elapsed until the electric charge transferred through a conductor reaches a given threshold value. 
For this purpose, we measure  the distribution of  the first-passage times for the net number of electrons 
transferred between two metallic islands in Coulomb blockade regime. 
Our experimental results are in excellent agreement with numerical calculations based on
a recent theory describing the exact first-passage-time distributions for  any  non-equilibrium stationary Markov process.
We also derive a simple analytical approximation for the first-passage-time distribution, 
which takes into account the non-Gaussian statistics of the electron transport, and 
show that it describes the experimental distributions with high accuracy.
This universal approximation describes a wide class of stochastic processes, and can be used beyond the context of mesoscopic charge transport.
In addition, we verify experimentally a fluctuation relation between the first-passage-time distributions 
for positive and negative thresholds.

\end{abstract}

\maketitle

\secs{Introduction} 
The {\em first-passage time} is  the  time it takes  a stochastic process 
 to first reach  a certain threshold. 
First-passage-time distributions have been studied in the context of Brownian motion~\cite{Smoluchowski,Schrodinger,Redner,Metzler},  
biochemistry~\cite{Szabo,Zwanzig,Galburt,Xie,Bauer,Benichou},  
astrophysics~\cite{Ch,Satya}, 
decision theory~\cite{Wald, Vergassola, Roldan},  
searching problems~\cite{Tejedor,Carlos},    
finance~\cite{Bouchaud,Perello}, and 
thermodynamics~\cite{SD,Ptaszynski,Roldan,PRX,Jordan,Garrahan,Hanggi}.
For example, in finance, the statistics of first passage times  is used in credit risk modelling, 
and in astrophysics, knowing the distribution of times required for a star in a globular cluster to
reach the escape velocity allows one to estimate the cluster's life-time \cite{Ch}. 
In the context of mesoscopic electron 
transport, the interest in the distributions of first passage times and  waiting times~\cite{Flindt1,Flindt2}
has been inspired by the tremendous progress in 
nanotechnology allowing very precise single-electron counting experiments \cite{Lu,Fujisawa,Gustavsson,Haug}.
Despite progress in the theory,  no experimental study of the statistics of the
time elapsed until the electric charge transferred through a conductor reaches a certain threshold,  
has been reported so far.   

The fluctuations  of a  stochastic process $N(t)$ are usually described in terms of the distribution $P_t(N)$ for the process to take the value $N$ at a fixed time $t$. An alternative approach is to study the first-passage-time probability distribution $\mathcal{P}_N(t)$ for a stochastic process to first reach or surpass a given value $N$ at time $t$. 
Recently, theories of the first-passage-time probability  
in Markovian systems have been developed~\cite{SD,Ptaszynski}.   
These methods provide the first-passage-time probabilities for the net number of jumps between 
any two states of the system. 
It has also been shown that first-passage-time distributions of currents~\cite{Jordan,Garrahan,Majumdar} 
and stopping-time distributions of entropy production~\cite{Roldan,PRX} 
satisfy universal laws for nonequilibrium steady-state obeying generalized detailed balance conditions. 
For these systems,  the  distributions  for the first time to produce  and 
reduce entropy by a certain amount 
have the same shape~\cite{Roldan,PRX}.  This {\em first-passage-time fluctuation relation}  
was generalized to stochastic processes describing the accumulation of evidence during sequential decision-making~\cite{Meik1,Meik2}.  
As a result, the first-passage-time distribution for one-dimensional biased random walk  
obeys, similarly to a Brownian particle~\cite{Smoluchowski,Schrodinger}, 
the relation $\mathcal{P}_N(t)=\mathcal{P}_{-N}(t) e^{vN/D}$, with $v$ and $D$ being the drift and diffusion coefficients in a lattice of unit spacing~\cite{Roldan}.

In this Letter, we report an experimental study 
of the first-passage-time statistics for electrons transferred 
through a metallic double dot in the Coulomb-blockade regime.
For this purpose, we obtain the full time-record of millions of
electron tunneling events between its two metallic islands (see Ref. \cite{Singh} for details). Subsequently, we compute the
first passage time distributions for the net number of electrons to reach a certain threshold.  
We find an excellent agreement with numerical calculations based on the exact theory~\cite{SD,Ptaszynski}.
We also derive and experimentally verify a simple and universal analytical expression for the distribution $\mathcal{P}_N(t)$.  
It depends on only three parameters -- the first three cumulants of the transferred charge distribution $P_t(N)$ -- 
and can be used to describe the first passage time fluctuations of a wide class of non-Gaussian stochastic processes not necessarily related to electronics.
Finally, we  test the first-passage-time fluctuation relation
closely related to the fluctuation theorem for the electron transport ~\cite{TN,Buttiker,Utsumi,Tasaki,FT1,FT2}.

\begin{figure}
	\includegraphics[width= 0.99 \columnwidth]{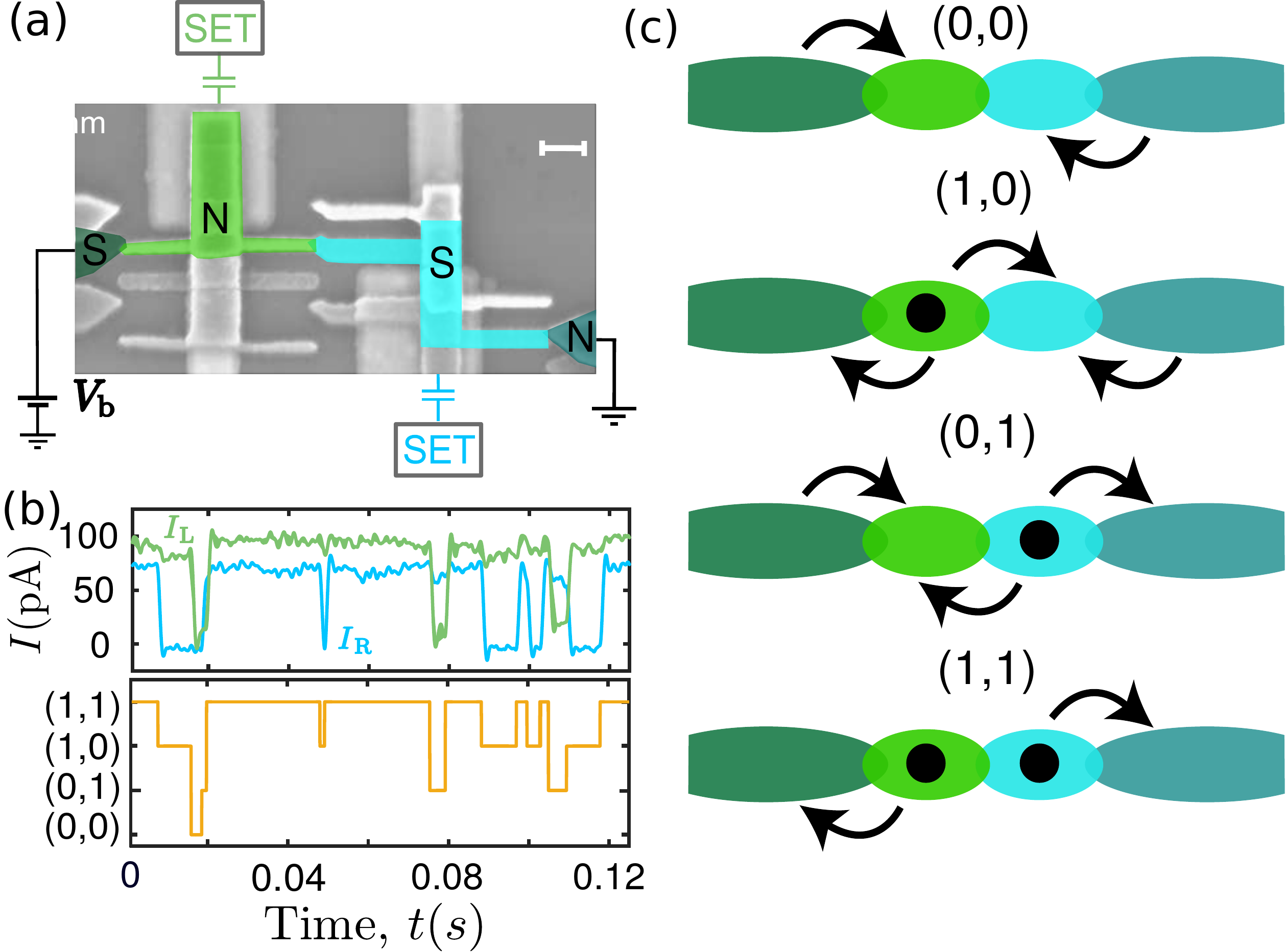}  
	\caption{\textbf{Experimental  setup.}  
		(a)
		Scanning electron micrograph of the double dot structure. 
		A DC bias voltage $V_{\rm b} = 90 \; \mu$V is applied to the sample. 
		One of the islands is made of normal metal (N, green), and the other one is superconducting (S, cyan).
		(b) Top panel: output currents of the detector SETs coupled to the left ($I_L$) and right ($I_R$) islands. 
         Bottom panel: time resolved trajectory of the charge states of the system $(N_L, N_R)$, with $N_{L,R} = 0$ or 1 
		indicating, respectively, the absence or the presence of an extra electron in the corresponding island.
		(c) A schematic sketch of the double dot. The arrows show all possible transitions in each charge state.
        The system is an example of an asymmetric simple exclusion process (ASEP) with two sites and open boundaries. 
	} 
	\label{fig:1} 
\end{figure}

\secs{Experiment} \label{Experiment}
Our metallic double dot contains aluminum superconducting parts together with normal
metal parts made of copper, see Fig. \ref{fig:1}(a). 
The left lead (dark green) and the right island (cyan) are superconducting,
while the left island (green)  and the right lead (turquoise)  are normal. 
Thus, all three tunnel junctions in our structure have a superconductor on one side
and a normal metal on the other. The double dot has a very high normal state resistance of $55$ M$\Omega$.
We run the experiment at the base temperature of 50 mK, where the two dots exhibit strong Coulomb blockade.
All these factors combined ensure very low tunneling rates of electrons, below 1 kHz.
We monitor the direction of electron jumps 
in and out of both islands using single-electron transistors~(SETs)
capacitively coupled to the  islands.

The top panel of Fig. \ref{fig:1}(b) shows an example of 
time traces of charge currents of the two SETs. 
We find that at chosen values of the gate potentials applied to the dots, 
and at bias voltage $V_{\rm b}=90$ $\mu$V  applied to the  device we can distinguish 
four populated charge states $(N_L,N_R)$, where $N_{L,R}=0$ or 1  
indicate the number of extra electrons in the islands. 
Monitoring the currents of both SETs, we detect the transitions between these charge states 
and find the corresponding transition rates. An example of a trajectory showing the transitions
between the charge states is shown in the bottom panel of Fig.~\ref{fig:1}(b).
Figure~\ref{fig:1}(c) shows a schematic sketch of all possible transitions 
from every charge state~\footnote{In our analysis, we do not consider cotunneling or Andreev tunneling due to high resistance of all junctions.}. 
The Markovian stochastic dynamics of the double dot is an example of an asymmetric simple exclusion process (ASEP)
with two sites and open boundaries~\cite{ASEP1,ASEP2,ASEP3,ASEP4,ASEP5}. 

\begin{figure} 
 	\includegraphics[width = \columnwidth]{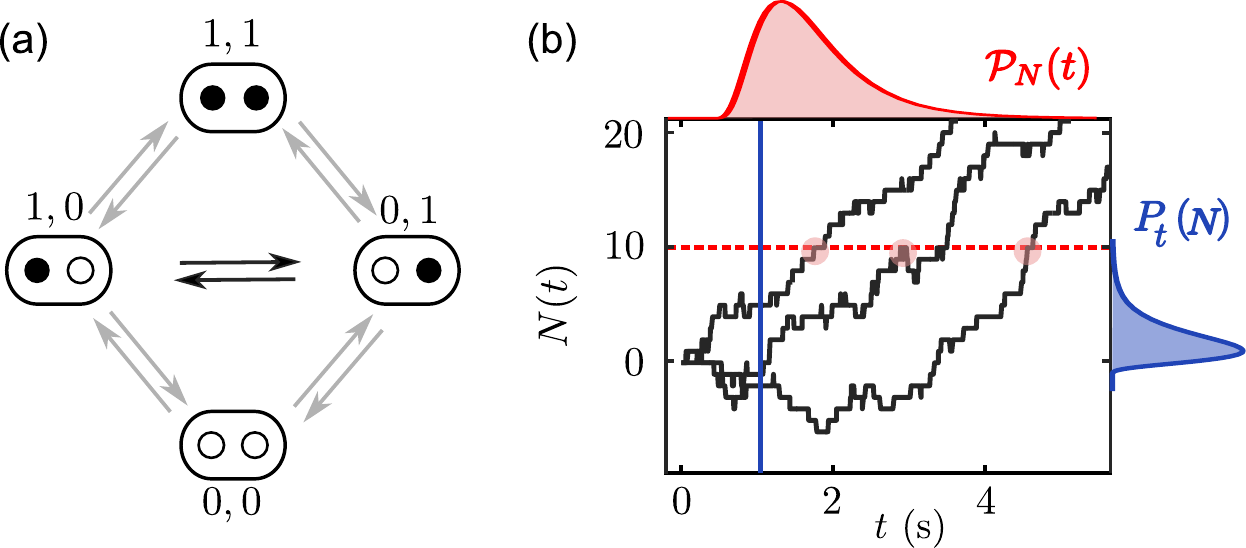} 
 	\caption{\textbf{Transition rates and first-passage-times.} 
		(a)
		Schematic depiction of the Markovian dynamics of the double dot system.  
		The transition rates $\Gamma_n^m$ from the state $n$ to the state $m$ have the following values:
        $\Gamma_{00}^{01}=644$ Hz, $\Gamma_{01}^{00}=131$ Hz,  
        $\Gamma_{00}^{10}=52$ Hz, $\Gamma_{10}^{00}=39$ Hz,
        $\Gamma_{01}^{11}=41$ Hz, $\Gamma_{11}^{01}=43$ Hz, $\Gamma_{10}^{11}=167$ Hz, $\Gamma_{11}^{10}=53$ Hz, $\Gamma_{01}^{10}=25$ Hz, $\Gamma_{10}^{01}=30$ Hz.
        We monitor the transitions between the states (1,0) and (0,1) shown by black arrows.
		(b)
		The solid black lines are sample time traces of the net number of electrons  transferred through the middle junction from the right to the left island $N(t)$. 
 		The horizontal dotted line indicates the fixed  threshold $N = 10$.
 		The first passage times for this threshold are marked with red circles, and we are interested in their distribution ${\cal P }_N(t)$ (red shaded area, illustration).
 		The vertical solid line indicates the time $t$ for which the distribution of the number of transferred electrons $P_t(N)$ is shown (blue shaded area, illustration). 
 	} 
 	\label{fig:2} 
 \end{figure}

\begin{figure*}
		\includegraphics[width= 2\columnwidth]{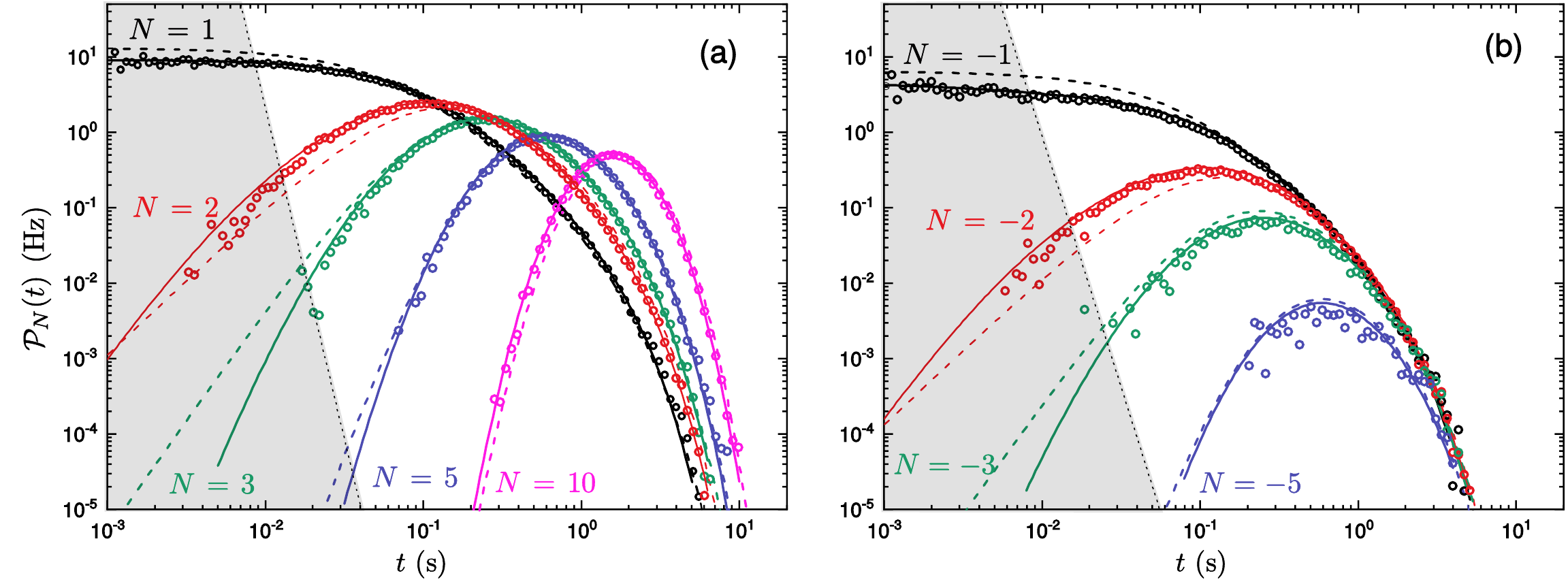} 
		\caption{\textbf{First-passage-time distribution} for positive (a) and negative (b) values of the threshold $N$.  
			Different colors correspond to different values of the threshold indicated in the figure.  		 
			Symbols are experimental data, solid lines --- numerics based on the exact theory \cite{SD},  		
			dashed lines --- Eq.~(\ref{PC3}). Shaded area approximately indicates the violation of the condition (\ref{cond2}).
			}
		\label{fig:fptd_t90_neffm1_10}
	\end{figure*}

Since we have full information about the population of the islands at any time, we can monitor the transition events between
all charge states.
Here we are interested in electron tunneling events through the middle junction from the right to the left island
indicated by black arrows in Fig.~\ref{fig:2}(a). 
In Fig.~\ref{fig:2}(b) we plot three example time traces of the net number of transmitted electrons $N(t)$. 
Next,  we look at the first-passage-times $t_i$ 
at which the traces $N(t)$ cross a chosen threshold $N$ for the first time.
Here $i$ enumerates different realizations of the experiment. 
The empirical distributions of these times for several values
of positive and negative thresholds, ${\cal P}_N(t)$, are shown in Fig.~\ref{fig:fptd_t90_neffm1_10}.
They constitute the main experimental result of this Letter.

\secs{Theory}
As a first step, we have numerically calculated~\cite{SM} first passage time distributions, 
which follow from the exact theory~\cite{SD,Ptaszynski}. We have used experimentally determined 
transition rates between the charge states of the double dot, given in the caption of Fig. \ref{fig:2}, as input parameters for the calculations.
The calculation results are shown by solid lines in Fig. \ref{fig:fptd_t90_neffm1_10}. We find perfect agreement between
the experiment and the numerical results, which confirms the consistency of our analysis. 

In addition to that, we propose and test a simple analytical expression for the first-passage-time distribution,
which takes into account the non-Gaussian statistics of the electron transport via a single parameter --- the third cumulant
of the distribution $P_t(N)$.
The cumulants of this distribution normalized by the observation time are defined as 
$ {\cal C}_n = \lim_{t\to \infty} t^{-1}(-i\partial/\partial\chi)^n\left.\ln\left[\sum_{N} e^{iN\chi} P_t(N)\right]\right|_{\chi=0}$.
In practice, the time $t$ should exceed the relaxation time of the system $\tau_r$ to ensure time independence of the measured cumulants, and
$\tau_r$ is defined as a time which the system needs to return back to the steady state after an external perturbation. 
The cumulants ${\cal C}_1$ and ${\cal C}_2$ are related to the average electric 
current $\langle I\rangle$ and the current noise $S_I=2\int dt \langle I(t)I(0) - \langle I\rangle^2 \rangle$
of the double dot as follows: $\langle I\rangle =e{\cal C}_1$ and $S_I = 2e^2{\cal C}_2$.
For a 1D biased random walk one finds ${\cal C}_1=v$ and ${\cal C}_2=2D$, with $v$ and $D$ being the drift and diffusion coefficients, respectively.

Our approximate expression for the first-passage-time distribution 
is based on the exact result for the one dimensional biased random walk~\cite{Feller,Redner},
${\cal P}_N(t)=|N|e^{-(\Gamma_++\Gamma_-)t}(\Gamma_+/\Gamma_-)^{N/2} I_N(2\sqrt{\Gamma_+\Gamma_-} t)/t$.
Here $\Gamma_\pm=({\cal C}_2\pm {\cal C}_1)/2$
are, respectively, the rates of jumping forward and backward.
This model also describes the transport of charged particles through a voltage biased tunnel junction~\cite{LLL}.
We adjust the three free parameters of the tunnel junction model, namely,
the rates $\Gamma_\pm$ and the effective particle charge $e^*$,  in such a way
that the first three cumulants of the charge transferred by particles in the model coincide
with the first three cumulants of the charge transferred by real electrons in the experiment~\cite{SM}. 
In particular, in this way we find $e^*=e\sqrt{{\cal C}_3/{\cal C}_1}$,
where $e$ is the electron charge. 
Afterwards, we approximate the exact first-passage-time distribution by the modified random walk expression
and arrive at our main theoretical result --- a simple analytical approximation for ${\cal P}_N(t)$,
\begin{eqnarray}
{\cal P}_N(t) &=& \frac{|N^*|e^{-\frac{{\cal C}_1{\cal C}_2}{{\cal C}_3}t}}{t} 
\left(\frac{{\cal C}_2+\sqrt{{\cal C}_1{\cal C}_3}}{{\cal C}_2-\sqrt{{\cal C}_1{\cal C}_3}}\right)^{\frac{N^*}{2}}
\nonumber\\ &&\times\,
I_{|N^*|}\left( \frac{{\cal C}_1\sqrt{{\cal C}_2^2-{\cal C}_1{\cal C}_3}}{{\cal C}_3}t \right). 
\label{PC3}
\end{eqnarray}
In this expression, $I_n(x)$ is the modified Bessel function of the first kind,   
and $ N^* = [N\sqrt{{{\cal C}_1}/{{\cal C}_3}}]$ is the threshold value for the number of virtual particles
such that the charge transmitted by them, $e^*N^*$, gets as close as possible 
to the net charge of real electrons $eN$.  
Here the square brackets $[\dots]$ denote the rounding function.
We have also assumed that ${\cal C}_1,{\cal C}_3>0$, and ${\cal C}_2^2>{\cal C}_1{\cal C}_3$. 
The approximate expression (\ref{PC3}) is valid if the conditions 
\begin{eqnarray}
\frac{{\cal C}_1|{\cal C}_1{\cal C}_4 - {\cal C}_2{\cal C}_3|}{12{\cal C}_2^3} \left( \frac{N}{{\cal C}_1t} -1\right)^2 \lesssim 1,\; t\gg \tau_r, \; |N|\gg 1
\label{cond2}
\end{eqnarray}
are fulfilled~\cite{SM}. The first condition implies that the approximation (\ref{PC3}) works in the vicinity 
of the maximum of the distribution ${\cal P}_N(t)$, occurring close to $t=N/{\cal C}_1$, but may fail  in the
tails of the distribution. At short times  the expression (\ref{PC3}) behaves as $t^{|N^*|-1}$. Provided the first and the third cumulants 
are not too far apart, $|{\cal C}_3-{\cal C}_1|\lesssim 0.5{\cal C}_1$, it reproduces  
the scaling of the exact distribution ${\cal P}_N(t)\sim t^{|N|-1}$ at small values of $N$.
In this case the last of the conditions (\ref{cond2}) may be relaxed.
In the long time limit  Eq. (\ref{PC3}) correctly reproduces the exponential decay of ${\cal P}_N(t)$ 
predicted by the exact theory~\cite{SD}, but may provide an inaccurate  decay rate if the first of the conditions (\ref{cond2}) is violated.
For a weakly non-Gaussian stochastic process with ${\cal C}_3,{\cal C}_4\lesssim {\cal C}_2$
the condition (\ref{cond2}) holds even at $t\to\infty$ and the approximation (\ref{PC3}) remains valid in this limit.    
In the Gaussian limit ${\cal C}_3\to 0$ the distribution (\ref{PC3})
reduces to the form  ${\cal P}_N(t) = {|N|\exp[-(N-{\cal C}_1t)^2/2{\cal C}_2t}]/{\sqrt{2\pi{\cal C}_2}t^{3/2}}$
well known from the theory of Brownian motion~\cite{Schrodinger,Smoluchowski},
while for ${\cal C}_3={\cal C}_1$ it returns to the original random walk form~\cite{Feller,Redner}.
We also note that the total probability of reaching a given threshold $A_N=\int_0^{\infty} dt\,\mathcal{P}_N(t)$ 
is equal to 1 for $N>0$ and is less than 1 for $N<0$.

The distribution (\ref{PC3}) satisfies the fluctuation relation for the first passage times
\begin{eqnarray}
\frac{{\cal P}_N(t)}{{\cal P}_{-N}(t)} = \frac{A_N}{A_{-N}} =
\left(\frac{{\cal C}_2+\sqrt{{\cal C}_1{\cal C}_3}}
{{\cal C}_2-\sqrt{{\cal C}_1{\cal C}_3}}\right)^{\left[N\sqrt{\frac{{\cal C}_1}{{\cal C}_3}}\right]}.
\label{FT}
\end{eqnarray}
This approximate relation does not require 
the system to be embedded in equilibrium environment or to exhibit detailed balance,
it only relies on the conditions (\ref{cond2}). 
In the limit $t\gg\tau_r$, and provided the system has a well defined temperature $T$,
one can prove an exact fluctuation relation \cite{SM}
\begin{eqnarray}
{{\cal P}_N(t)}/{{\cal P}_{-N}(t)} = \exp[N eV_{\rm b}/k_BT],
\label{FTT}
\end{eqnarray}
which is consequence of the fluctuation theorem for the electron transport~\cite{TN,Buttiker,Utsumi,Tasaki}.
The relations (\ref{FT}) and (\ref{FTT}) are close to each other in the common range of validity.
They become equivalent, for example, for a Gaussian equilibrium stochastic process describing charge transport through an Ohmic resistor, 
in which case ${\cal C}_2=2k_BT{\cal C}_1/eV_{\rm b}$ and ${\cal C}_3\to 0$,
and for a biased tunnel junction, for which ${\cal C}_3={\cal C}_1$ and ${\cal C}_2={\cal C}_1\coth[{eV_{\rm b}}/{2k_BT}]$.

 \begin{figure}
 	\includegraphics[width= 0.88 \columnwidth]{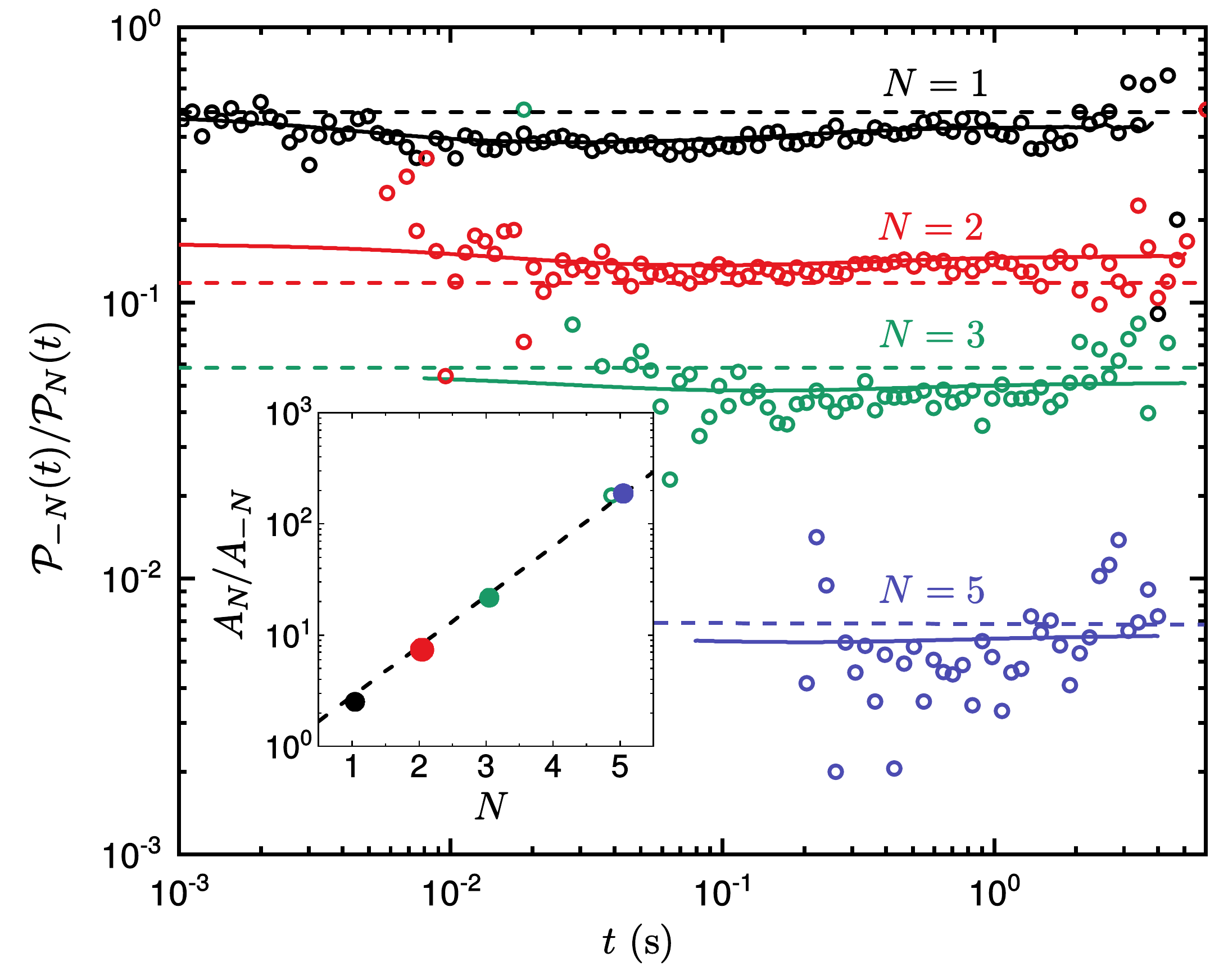} 
 	\caption{\textbf{First-passage-time fluctuation relation.} 
        The ratio ${{\cal P}_N(t)}/{{\cal P}_{-N}(t)}$ for several values of $N$.
 		Symbols are  the experimental data, solid lines ---  exact theory~\cite{SD}, dashed lines -- approximation (\ref{FT}).
        Inset: the ratio of the total probabilities integrated over time, $A_{N}/A_{-N}$; symbols are the experimental data,
        dashed line -- approximation (\ref{FT}).
}
 	\label{fig:fptd_fluc}
 \end{figure}

\secs{Discussion} 
We have determined the transition rates between the charge states of the double dot, 
given in the caption of Fig. \ref{fig:2}, in the standard way 
by counting the number of corresponding transitions per second and normalizing the result by state occupation probabilities. 
The numerical calculations based on the exact theory~\cite{SD,Ptaszynski} 
with independently determined rates
agrees well with the experimental distributions, see Fig. \ref{fig:fptd_t90_neffm1_10}.


Next, we test the approximate expression (\ref{PC3}).
Having determined the rates,  we have used the full counting statistics formalism~\cite{BN}  and found the first four cumulants of the charge distribution,
${\cal C}_1=4.60$ Hz, ${\cal C}_2=9.27$ Hz,  ${\cal C}_3=2.18$ Hz,  ${\cal C}_4=3.96$ Hz.  
As a consistency check, 
we have also determined the cumulants directly from the measured distributions of the number of transmitted electrons $P_t(N)$, and 
obtained the same values within $\pm 0.2$ Hz, which is compatible with statistical uncertainty.
The system relaxation time is given by the inverse of the eigenvalue of the transition rates matrix,
which has the real part closest to zero among its non-zero eigenvalues,  and equals $\tau_r=8.7$ ms~\footnote{this time scale is larger than the electron-phonon relaxation time $\sim 10 ^{-6}$~s, and electron-electron relaxation time $\sim 10 ^{-9}$~s.}.
With these values of the cumulants the first of the conditions (\ref{cond2}) is fulfilled at $t\to\infty$
and the expression (\ref{PC3}) fits the experimental data very well in the long time limit, see Fig. \ref{fig:fptd_t90_neffm1_10}.
The first two of the conditions (\ref{cond2}) are violated in the shaded areas of Figs. \ref{fig:fptd_t90_neffm1_10}(a,b).  
We find that Eq. (\ref{PC3}) fits the experimental data rather well outside these areas
even for small values of the threshold $|N|=1,2$, 
which is explained by the relatively small difference between the cumulants ${\cal C}_1$ and ${\cal C}_3$. 
We have found that at another value of the bias voltage, at which the difference between  ${\cal C}_1$ and ${\cal C}_3$
is bigger, Eq. (\ref{PC3}) has worked only for sufficiently large $|N|$ \cite{SM}.

Finally, we have tested the fluctuation relation (\ref{FT}) by comparing it with
the experimental data and with the numerics based on the full theory \cite{SD,Ptaszynski}.
The result of this comparison is shown in Fig. \ref{fig:fptd_fluc}. 
We have again found that the numerics provided a very accurate match with the data.
The approximation (\ref{FT}), although less accurate, also describes the experiment rather well. 
The exact numerical analysis reveals the approximate nature of the fluctuation relations (\ref{FT},\ref{FTT}) for non-equilibrium systems
with broken detailed balance, like our double dot. Indeed, the solid lines in Fig. \ref{fig:fptd_fluc}, showing the exact results, 
slightly deviate from constant values.

\secs{Conclusion} 
We have measured the distribution of the first passage times for electrons 
tunneling between two islands in the   Coulomb blockade regime employing single electron counting technique.
We have compared the experimental results with the predictions of the exact theory~\cite{SD,Ptaszynski}
and observed perfect agreement. Besides that, we have proposed a simple approximation for the distribution of the first passage times (\ref{PC3}), 
which accounts for the non-Gaussian statistics of single-electron tunneling via the third cumulant of the distribution 
of the number of transmitted electrons. This universal result should be applicable to any stochastic process
provided the conditions (\ref{cond2}) are satisfied.
We have demonstrated that the expression (\ref{PC3}) matches the experimental data quantitatively without any free parameter at sufficiently long times determined by (\ref{cond2}). 
Finally, we have experimentally verified
a fundamentally important fluctuation relation for the distribution of the first passage times~(\ref{FT}).

We acknowledge the provision of facilities by Aalto University 
at OtaNano Micronova Nanofabrication Centre
and the computational resources provided by the Aalto Science-IT project.
We thank Matthias Gramich and Libin Wang for technical assistance.
We thank Abhishek Dhar and Izaak Neri for fruitful discussions.
This work is partially supported
by Academy of Finland, Project Nos. 284594, 272218, and 275167~(S.~S., D.~S.~G., J.~T.~P., and J.~P.~ P.),
by European Research Council~(ERC) under the European Union’s Horizon 2020 research
and innovation programme under grant agreement No. 742559 (SQH),
by the Russian Foundation for Basic Research and German Research Foundation (DFG) Grant No. KH~425/1-1 (I.~M.~K.), and  
by JSPS Grants-in-Aid for Scientific Research (JP16H02211 and JP17K05587)~(K.~S.).
Correspondence and requests for materials should be addressed to S.~S.~(email: \url{sshilpi916@gmail.com}).

%

\newpage

\appendix 	

%

\section{Exact Theory}
\label{exact_theory}

Here we briefly describe how to derive the exact expression for the first passage time distribution in a system described by a master equation following Refs.~\cite{SD} and \cite{Ptaszynski} and comment on how to calculate it in practice.
The system state is fully described by the vector $\bm p$ containing the occupation probabilities $p_j$ of the four charge states of the double dot.
In matrix form, the master equation describing its time evolution can be written as
\begin{equation}
\dot{\bm p}=\hat{\bm W}_0\, {\bm p},
\end{equation}
where the elements of the rate matrix $\hat{\bm W}_0$ are $(W_0)_{k}^j=\Gamma_{k}^j - \delta_{jk}\sum_{k' \not= j}\Gamma_{j}^{k'}$ and $\Gamma_{k}^j$ is the transition rate from the state $k$ to the state $j$.

The fluctuations of the number of electrons transferred through the middle junction are described by the full counting statistics formalism ~\cite{BN}.
Following it, we introduce a counting parameter $z$ and a new rate matrix $\hat{\bm W}(z)$.
The latter differs from $\hat{\bm W}_0$ only in two entries, namely, $W_{01}^{10}(z) = z\, \Gamma_{01}^{10}$ and $W^{01}_{10}(z) = z^{-1}\, \Gamma^{01}_{10}$.
Next, we introduce two further quantities.
The first is the transition matrix $\hat{\bm T}(N,t)$ in extended state space.
Its entries $T(N,t)_k^j$ are the probabilities to go from state $k$ to state $j$ in time $t$ while increasing the net number of counted transitions by $N$.
In the full counting statistics framework, it is given by the expression
\begin{equation} \label{eq:Tnt}
\hat{\bm T}(N,t) = \frac{1}{2\pi i} \oint \frac{dz}{z^{N+1}} \exp(\hat{\bm W}(z) t) .
\end{equation}
Second, we define a matrix $\hat{\bm F}(N, t)$ so that the entries $F(N,t)_k^j\, dt$ are the probabilities to go from state $k$ to state $j$ in time $t$ while hitting the threshold $N$ for the first time during the time interval $[t, t + dt]$.
These two quantities are connected by the intuitive relation
\begin{equation} \label{eq:TandF}
\hat{\bm T}(N,t) = \int_0^t d\tau\, \hat{\bm T}(0, t-\tau)\, \hat{\bm F}(N,\tau) .
\end{equation}

The first passage time distribution can be obtained from ${\cal P}_N(t) = \operatorname{tr}( \hat{\bm F}(N,t) {\bm p}_{\text{st}} )$, where ${\bm p}_{\rm st}$ is the stationary solution of the master equation and the trace of a vector is defined as the sum of its entries.
We solve Eq.~(\ref{eq:TandF}) for $\hat{\bm F}$ by taking the Laplace transform, obtaining the final result
\begin{equation}
{\cal P}_N(t) = \int_{c-i\infty}^{c+i\infty} \frac{ds}{2\pi i}\, e^{st} \operatorname{tr}\left[ \hat{\bm T}(0,s)^{-1} \hat{\bm T}(N,s) {\bm p}_{\mathrm{st}} \right] .
\label{P_exact}
\end{equation}
Here $c$ is a positive real number with dimension of frequency.
Eq.~(\ref{P_exact}) is the basis of our exact numerical results.

Next, we discuss how to calculate the Laplace transform of the matrix (\ref{eq:Tnt}),
\begin{equation}
\hat{\bm T}(N,s) = \int_0^\infty \frac{dt}{2\pi i} \oint \frac{dz}{z^{N+1}} \exp([\hat{\bm W}(z) - s]\, t) .
\end{equation}
Note that the value of this expression does not depend on the choice of contour encircling $z=0$.
We want to exchange the order of the integrals and thus have to choose the contour in such a way that the real parts of all eigenvalues of $\hat{\bm W}(z) - s$ are negative along the contour.
For all $s$ with positive real part, this can be ensured by using the unit circle as the integration contour, since all eigenvalues of $\hat{\bm W}(z)$ have a non-positive real part for $|z| = 1$ in our system.

In order to evaluate $\hat{\bm T}(N,s)$ in practice, we note the fact \cite{SD} that the polynomial $z \det(\hat{\bm W}(z) - s)$ has two zeros $z_\pm(s)$ satisfying $| z_+(s) | \geq 1 > | z_-(s) |$.
According to our discussion above, $|z_+(s)|$ can not be equal to unity for $s$ with positive real part and we can write
\begin{align}
	\hat{\bm T}(N,s) &= \frac{1}{2\pi i}\oint \frac{dz}{z^{N+1}} \left[ s - \hat{\bm W}(z) \right]^{-1} \\
	&= \operatorname{Res}_0\Big( \frac{[ s - \hat{\bm W}(z) ]^{-1}}{z^{N+1}} \Big) \notag \\
	&\qquad	+ \operatorname{Res}_{z_-(s)} \Big( \frac{[ s - \hat{\bm W}(z) ]^{-1}}{z^{N+1}} \Big) ,
\end{align}
where the integration contour now is the unit circle and $\operatorname{Res}_{z_0}(f(z))$ denotes the residue of $f(z)$ at $z \to z_0$.

In general, it is not possible to express the matrix $\hat{\bm T}(N,s)$ in terms of elementary functions, and the Fourier integral (\ref{P_exact}) has to be performed numerically.
To this end, we sample the integrand
\begin{equation}
I_c(x) = \operatorname{tr}\left[ \hat{\bm T}(0,c + ix)^{-1} \hat{\bm T}(N,c + ix) {\bm p}_{\mathrm{st}} \right]
\end{equation}
of the expression ${\cal P}_N(t) = \frac{e^{ct}}{2\pi} \int_{-\infty}^\infty e^{ixt} I_c(x)\, dx$ at values $x = 0, \Delta x, 2\Delta x, \dots, x_{\mathrm{max}}$ (note that $I_c(-x) = I_c(x)^\ast$).
By the Shannon-Nyquist theorem, it can be expected that the maximum time where the numerical integration yields reliable results is of the order
\begin{equation}
t_{\text{max}} \sim \frac 1 c\, W\!\left( \frac{c\pi}{\Delta x} \right) ,
\end{equation}
where the Lambert $W$ function was added to compensate for the $e^{ct}$ factor.
Similarly, the minimum time is of the order $t_{\text{min}} \sim \frac 1 c\, W\!\big( \frac{c\pi}{x_{\text{max}}} \big)$.

\section{Derivation of Eq. (\ref{PC3})} 
\label{derivation}

Following the method of the full counting statistics~\cite{BN}, we introduce  the cumulant generating
function (CGF)  ${\cal F}(\chi)$, which is related to the distribution $P_t(N)$ as
\begin{eqnarray}
e^{{\cal F}(\chi) t} = \sum_N e^{i N\chi} P_t(N),\;\; t\gg \tau_r.
\end{eqnarray} 
Accordingly, the cumulants can be expressed as
\begin{eqnarray}
{\cal C}_n = \left(-i{\partial}/{\partial\chi}\right)^n{\cal F}(\chi)\big|_{\chi=0}.
\label{Cn}
\end{eqnarray}
As a consequence, Taylor expansion of ${\cal F}(\chi)$ has the form
\begin{eqnarray}
{\cal F}(\chi) = i{\cal C}_1\chi - \frac{{\cal C}_2}{2}\chi^2 - i \frac{{\cal C}_3}{6}\chi^3 + \frac{{\cal C}_4}{24}\chi^4 + \dots.
\label{FCS}
\end{eqnarray}

In the long time limit $t\gg \tau_r$ the number of electrons transmitted
through the double dot $N(t)$ can be approximately treated as a Markov process. In this case
the distribution of the number of transmitted electrons and
first passage time distribution are related as~\cite{Siegert} 
\begin{eqnarray}
P_t(N) = \int_0^t dt' {\cal P}_{N}(t') P_{t-t'}(0).
\label{Eq_Siegert}
\end{eqnarray} 
This equation is a simplified, one-dimensional, version of the matrix equation (\ref{eq:TandF}).
Taking the Laplace transform of it, we arrive at the explicit expression for the first passage time distribution
in the form
\begin{eqnarray}
{\cal P}_N(t) = \int_{c-i\infty}^{c+i\infty} \frac{ds}{2\pi i} e^{st} 
\frac{\int_{-\pi}^\pi \frac{d\chi}{2\pi} \frac{e^{-i\chi N}}{s-{\cal F}(\chi)}}
{\int_{-\pi}^\pi \frac{d\chi}{2\pi} \frac{1}{s-{\cal F}(\chi)}}.
\label{cal_P}
\end{eqnarray}

In the simplest Gaussian approximation one can keep only the first two  terms of the Taylor expansion (\ref{FCS}). In  this case 
the integral (\ref{cal_P}) can be taken exactly, which results in the distribution of first passage times of
a Brownian particle ~\cite{Schrodinger,Smoluchowski}
\begin{eqnarray}
{\cal P}_N(t) = \frac{|N|e^{-(N-{\cal C}_1t)^2/2{\cal C}_2t}}{\sqrt{2\pi{\cal C}_2}t^{3/2}}.
\label{P_diff}
\end{eqnarray}
This approximation is valid if (see the next section)
\begin{eqnarray}
\frac{{\cal C}_3{\cal C}_1}{3{\cal C}_2^2}\left|\frac{N}{{\cal C}_1t}-1\right|\ll 1.
\label{cond2_app}
\end{eqnarray}
Our goal is to improve this approximation by taking into account the third order contribution of the Taylor expansion (\ref{FCS}). 
Unfortunately, if one directly substitutes the cubic polynomial containing the terms up to $\propto\chi^3$ of 
the CGF (\ref{FCS}) into the expression (\ref{cal_P}), the resulting integral cannot be
evaluated analytically in a simple form. The way around this problem is to choose an alternative approximate form of the 
cumulant generating function, ${\cal F}_{\rm app}(\chi)$, such that its Taylor expansion in $\chi$ would give 
the same first three cumulants as the original generating function ${\cal F}(\chi)$ and, at the same time,
the integral (\ref{cal_P}) would be analytically solvable.
We achieve this by choosing
\begin{eqnarray}
{\cal F}_{\rm app}(\chi)=\Gamma_+\left( e^{i\alpha\chi} -1 \right) + \Gamma_-\left( e^{-i\alpha\chi} -1 \right), 
\label{Fapp}
\end{eqnarray}
where
\begin{eqnarray}
\Gamma_\pm = \frac{{\cal C}_1\left({\cal C}_2 \pm \sqrt{{\cal C}_1{\cal C}_3}\right)}{2{\cal C}_3},\;\;
\alpha=\sqrt{\frac{{\cal C}_3}{{\cal C}_1}}.
\label{Gamma_pm}
\end{eqnarray}
For simplicity, here we have assumed that ${\cal C}_1$ and ${\cal C}_3$ have the same sign and the parameter $\alpha$ is real valued.
One can easily verify that the first three terms of the Taylor expansion of the CGF (\ref{Gamma_pm}) coincide with the first three 
terms of the expansion (\ref{FCS}). 
The generating function (\ref{Fapp}) describes, for example, the process of random bi-directional  Poissonian tunneling of charged particles
with the effective electric charge $e^*=\alpha e$ through a biased tunnel junction ~\cite{LLL}, 
in which the rate of tunneling  forward, $\Gamma_+$, differs form the backward rate $\Gamma_-$. 
The same generating function describes random walk along a one-dimensional chain of sites separated by distance $\alpha$ ~\cite{Feller,Redner}.
Since the period of the CGF is different from $2\pi$ due to the change of the effective particle charge,  
we should adjust the the integration limits in the Eq. (\ref{cal_P}) and write it in the form 
\begin{eqnarray}
{\cal P}_{N}^{\rm app}(t)=\int\frac{ds}{2\pi i}\, e^{s t} \,
\frac{\int_{-\pi/\alpha}^{\pi/\alpha} \frac{d\chi}{2\pi}  \frac{e^{-i\chi\alpha N^*}}{s - {\cal F}_{\rm app}(\chi)}}
{\int_{-\pi/\alpha}^{\pi/\alpha} \frac{d\chi}{2\pi}  \frac{1}{s - {\cal F}_{\rm app}(\chi)}}.
\label{cal_P_app}
\end{eqnarray}  
Here the threshold $N^*$ for the effective, or virtual, particle is related to the original threshold value for the number of electrons, $N$, as follows
\begin{eqnarray}
N^* = \left[ {N}/{\alpha} \right].
\label{Nstar}
\end{eqnarray}
$N^*$ is chosen in such a way that the charge transmitted by virtual particles, $e^*N^*$, approaches the charge transferred by electrons, $eN$, 
as close as possible.

As a first step, we evaluate the integral
\begin{eqnarray}
&& \int_{-\pi/\alpha}^{\pi/\alpha} \frac{d\chi}{2\pi}  \frac{e^{-i\chi\alpha n}}{s - {\cal F}_{\rm app}(\chi)} =
\frac{1}{\alpha \sqrt{(s+\Gamma_\Sigma)^2 - 4\Gamma_+\Gamma_-}}
\nonumber\\ &&\times\,
\frac{(2\Gamma_+)^n}{\left(s+\Gamma_\Sigma + \sqrt{(s+\Gamma_\Sigma)^2 - 4\Gamma_+\Gamma_-} \right)^n},
\end{eqnarray}
where $\Gamma_\Sigma=\Gamma_++\Gamma_-$ and $n$ is any integer number. 
Substituting this expression in the Eq. (\ref{cal_P_app}), we find 
\begin{eqnarray}
{\cal P}_{N}^{\rm app}(t)=\int\frac{ds}{2\pi i}
\frac{(2\Gamma_+)^{N^*} e^{s t}}{\left(s+\Gamma_\Sigma + \sqrt{(s+\Gamma_\Sigma)^2 - 4\Gamma_+\Gamma_-} \right)^{N^*}}.
\nonumber
\end{eqnarray} 
This integral can be analytically solved,  which gives the well known result~\cite{Feller,Redner}
\begin{eqnarray}
{\cal P}_{N}^{\rm app}(t)=e^{-\Gamma_\Sigma t} \frac{|N^*|}{t}\left(\frac{\Gamma_+}{\Gamma_-}\right)^{\frac{N^*}{2}} I_{N^*}\left(2\sqrt{\Gamma_+\Gamma_-}t\right).
\nonumber\\
\label{Papp_final}
\end{eqnarray} 
This expression, in combination with Eqs. (\ref{Gamma_pm}) and (\ref{Nstar}), takes the form given in the main text,
\begin{eqnarray}
{\cal P}_N(t) &=& \frac{|N^*|e^{-\frac{{\cal C}_1{\cal C}_2}{{\cal C}_3}t}}{t} 
\left(\frac{{\cal C}_2+\sqrt{{\cal C}_1{\cal C}_3}}{{\cal C}_2-\sqrt{{\cal C}_1{\cal C}_3}}\right)^{\frac{N^*}{2}}
\nonumber\\ &&\times\,
I_{|N^*|}\left( \frac{{\cal C}_1\sqrt{{\cal C}_2^2-{\cal C}_1{\cal C}_3}}{{\cal C}_3}t \right). 
\label{PC3_app}
\end{eqnarray}

It is interesting that in the limit $[{{\cal C}_1\sqrt{{\cal C}_2^2-{\cal C}_1{\cal C}_3}}/{{\cal C}_3}]t\gtrsim 1$ 
with good accuracy one can approximate the Eq. (\ref{PC3}) by the diffusion formula (\ref{P_diff}) with renormalized cumulants
\begin{eqnarray}
{\cal C}_1 &\to & \tilde{\cal C}_1 
= {\cal C}_1\sqrt{\frac{2\sqrt{{\cal C}_2^2 - {\cal C}_1{\cal C}_3}}{{\cal C}_2+\sqrt{{\cal C}_2^2 - {\cal C}_1{\cal C}_3}}},
\nonumber\\
{\cal C}_2 &\to & \tilde{\cal C}_2 = \sqrt{{\cal C}_2^2 - {\cal C}_1{\cal C}_3}.
\label{tilde_C12}
\end{eqnarray} 
Accordingly, in the long time limit the first passage time distribution acquires the universal form ~\cite{SD} 
\begin{eqnarray}
{\cal P}_N(t) \to \frac{A(N)}{t^{3/2}} e^{-\tilde\Gamma t}
\label{P_SD}
\end{eqnarray}  
with
\begin{eqnarray}
\tilde\Gamma = \frac{\tilde{\cal C}_1^2}{2\tilde{\cal C}_2}=\frac{{\cal C}_1^2}{{\cal C}_2+\sqrt{{\cal C}_2^2 - {\cal C}_1{\cal C}_3}}.
\end{eqnarray}

\section{Fluctuation relation for the first passage time distribution}

The fluctuation relation (\ref{FT}) directly follows from the approximate expression for the first passage time distribution (\ref{PC3}).
One might wonder if this relation remains valid if one goes beyond that approximation. Here we will show that in the long time limit $t\gg\tau_r$
and in equilibrium the relation (\ref{FTT}) in Main  Text follows from the fluctuation theorem for the electron transport. The latter theorem states that at $t\gg\tau_r$
the distribution of transmitted charges has the property
\begin{eqnarray}
\frac{P_t(N)}{P_t(-N)} = e^{eV_{\rm b}/k_BT},
\label{FT2}
\end{eqnarray} 
where $T$ is the temperature of the leads and of the electro-magnetic environment. It is known that the theorem (\ref{FT2}) holds for virtually any
mesoscopic system in thermal equilibrium. Combining Eqs. (\ref{FT2}) and (\ref{Eq_Siegert}) one can easily see that the fluctuation relation
\begin{eqnarray}
\frac{{\cal P}_N(t)}{{\cal P}_{-N}(t)} = e^{eV_{\rm b}/k_BT}
\label{FT3}
\end{eqnarray}  
should hold. The Eq. (\ref{FT}) given in the main text is more general in the sense that it remains valid out of equilibrium.
On the other hand, Eq. (\ref{FT3}) is not restricted by the times at which only first three cumulants are relevant, and hence
it is more accurate then the relation (\ref{FT}) if thermal equilibrium in the leads is maintained at high bias.

\section{Validity of the approximations}
\label{conditions}

In order to derive the validity conditions for the approximations (\ref{P_diff}) and (\ref{PC3_app}), we assume that the time $t$ is sufficiently long 
and the threshold value is sufficiently large, $|N|\gg 1$. In this case we can use saddle point approximation while evaluating the integrals.
We begin with the approximate expression (\ref{cal_P}), in which  we make the following approximation
\begin{eqnarray}
\frac{\int_{-\pi}^\pi \frac{d\chi}{2\pi} \frac{e^{-i\chi N}}{s-{\cal F}(\chi)}}
{\int_{-\pi}^\pi \frac{d\chi}{2\pi} \frac{1}{s-{\cal F}(\chi)}} \approx 
e^{-i\chi_p(s) N}.
\end{eqnarray}
Here $\chi_p(s)$ is the solution of the equation 
$
{\cal F}(\chi_p)=s
$
and defines the position of a pole 
of the function $1/[s-{\cal F}(z)]$ in the complex plain of the parameter $z=e^{i\chi}$. 
After that, the first passage time distribution (\ref{cal_P}) acquires the form
\begin{eqnarray}
{\cal P}_N(t) = \int_{c-i\infty}^{c+i\infty} \frac{ds}{2\pi i} e^{st} e^{-i\chi_p(s) N}.
\end{eqnarray}
We apply saddle point approximation for this integral, which leads to the result
\begin{eqnarray}
{\cal P}_N(t) \approx A e^{s_st} e^{-i\chi_p(s_s) N}.
\label{Papp}
\end{eqnarray}
Here $A$ is a certain pre-factor weakly dependent on time, and $s_s$ is the solution of the saddle point equation
\begin{eqnarray}
t = iN[{d\chi_p(s_s)}/{ds}]. 
\end{eqnarray}
Combining this equation with the condition ${\cal F}(\chi_p)=s$, we can exclude the saddle point value $s_s$ and express the result (\ref{Papp}) as
\begin{eqnarray}
{\cal P}_N(t) \approx A e^{{\cal F}(\chi_p)t} e^{-i\chi_p N},
\label{Papp1}
\end{eqnarray}
where $\chi_p$ should be found from the equation
\begin{eqnarray}
{\cal F}'(\chi_p) = {iN}/{t}.
\label{saddle_point}
\end{eqnarray}

If it is possible to express the exact CGF as the sum  
of CGF of an exactly
solvable model, ${\cal F}_0(\chi)$,  and a small correction to it $\delta{\cal F}(\chi)$, 
then, to the lowest order in the correction, 
the equation (\ref{saddle_point}) takes the form
\begin{eqnarray}
{\cal F}'_0(\chi_{0}) + {\cal F}''_0(\chi_{0})\delta\chi_p + \delta{\cal F}'(\chi_{0})  = {iN}/{t}.
\label{sp}
\end{eqnarray}
Here $\chi_{0}$ is the solution of the equation 
\begin{eqnarray}
{\cal F}'_0(\chi_{0})={iN}/{t},
\label{saddle_app}
\end{eqnarray} 
and $\delta\chi_p$ is the small correction. 
The solution of the Eq. (\ref{sp}) is $\delta\chi_p = -\delta{\cal F}'(\chi_{0})/{\cal F}''_0(\chi_{0}) $.
Substituting this result back in the Eq. (\ref{Papp1}) and performing its expansion up to the lowest order in $\delta{\cal F}$, we find
\begin{eqnarray}
{\cal P}_N(t) \approx A e^{{\cal F}_0(\chi_{0})t-i\chi_{0} N} e^{\delta{\cal F}(\chi_{0})t}.
\label{Papp2}
\end{eqnarray}
Comparing this expession with the Eq. (\ref{Papp1}),
we conclude that the exact CGF ${\cal F}(\chi)$ can be replaced by the approximate one, ${\cal F}_0(\chi)$, as long as
\begin{eqnarray}
|\delta{\cal F}(\chi_{0})|t\ll \max\{{\cal F}_0(\chi_{0})t-i\chi_{0} N\}-{\cal F}_0(\chi_{0})t+i\chi_{0} N.
\nonumber\\
\label{cond3}
\end{eqnarray}

\begin{figure*}[!ht]
	\includegraphics[width= 1.5 \columnwidth]{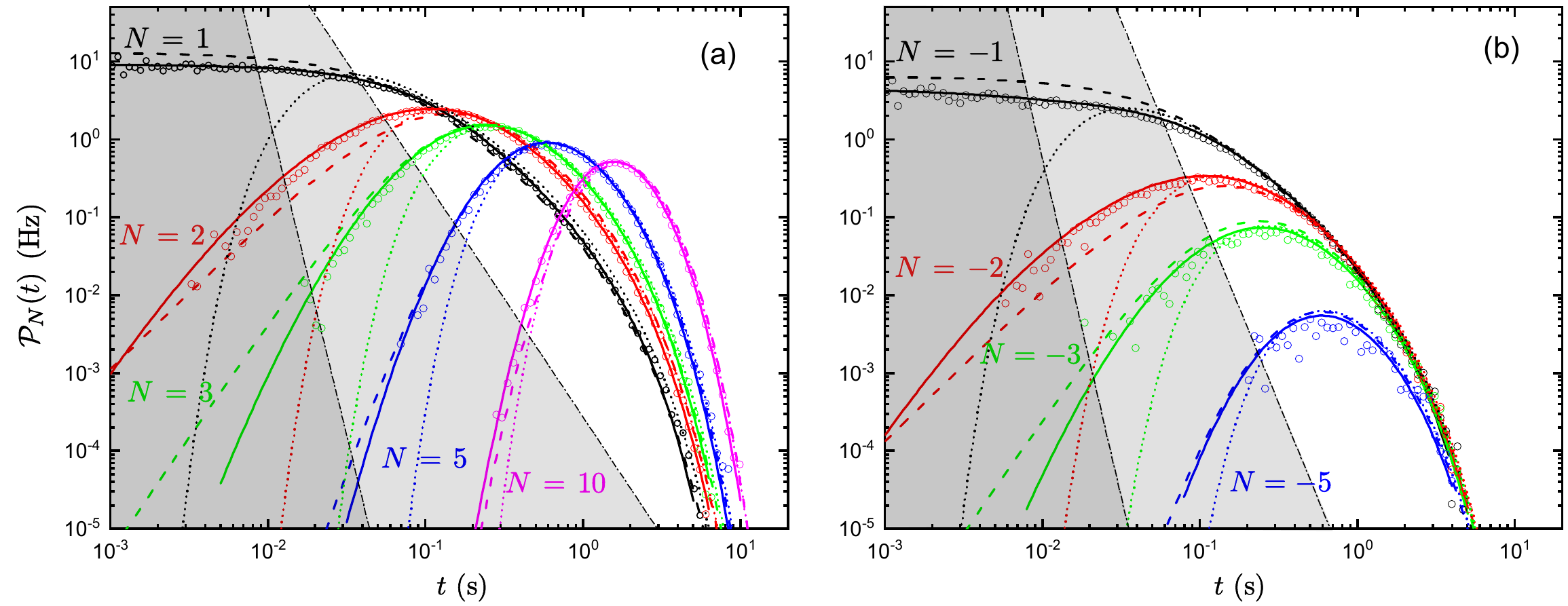}
	\caption{\textbf{First passage time distribution} at bias voltage $V_{\rm b}=90$ $\mu$V
        for diferent values of the threshold $N$  in the direction of the flow (a) and  against the flow (b).
		The symbols are the experimental data, solid lines show full numerical theory (\ref{P_exact}),  		
		dashed lines are the predictions of the Eq. (\ref{PC3_app}), and dotted lines ---  the diffusion, or Gaussian, approximation (\ref{P_diff}).
        Left dashed-dotted black lines mark the short time validity condition~\eqref{cond3_app} of the non-Gaussian approximation (\ref{PC3_app}),
        and the right dashed-dotted lines --- the validity boundary (\ref{cond2_app}) of the Gaussian approximation (\ref{P_diff}).
        The transition rates for this bias volatge are $\Gamma_n^m$ from the state $n$ to the state $m$ have the following values:
        $\Gamma_{00}^{01}=644$ Hz, $\Gamma_{01}^{00}=131$ Hz,  
        $\Gamma_{00}^{10}=52$ Hz, $\Gamma_{10}^{00}=39$ Hz,
        $\Gamma_{01}^{11}=41$ Hz, $\Gamma_{11}^{01}=43$ Hz, $\Gamma_{10}^{11}=167$ Hz, $\Gamma_{11}^{10}=53$ Hz, $\Gamma_{01}^{10}=25$ Hz, $\Gamma_{10}^{01}=30$ Hz.
        The cumulants have the following values ${\cal C}_1=4.60$ Hz, ${\cal C}_2=9.27$ Hz,  ${\cal C}_3=2.18$ Hz,  ${\cal C}_4=3.96$ Hz.
}
	\label{fig:fptd_comp}
\end{figure*}

Gaussian approximation for the first passage time distribution (\ref{P_diff}) results from the  Gaussian approximation for the CGF
\begin{eqnarray}
{\cal F}_0(\chi) = i{\cal C}_1\chi - {\cal C}_2{\chi^2}/{2}. 
\end{eqnarray}
In this case the solution of the approximate saddle point equation (\ref{saddle_app}) reads
\begin{eqnarray}
\chi_0=\frac{{\cal C}_1}{i{\cal C}_2}\left(\frac{N}{{\cal C}_1t}-1\right).
\end{eqnarray}
The correction to the CGF in this case is estimated as $\delta F(\chi)={\cal C}_3(i\chi)^3/6$, and hence the condition (\ref{cond3}) acquires the form 
\begin{eqnarray}
\frac{{\cal C}_3{\cal C}_1^3}{6{\cal C}_2^3}\left|\frac{N}{{\cal C}_1t}-1\right|^3 t\ll \frac{{\cal C}_1^2}{2{\cal C}_2}\left|\frac{N}{{\cal C}_1t}-1\right|^2t.
\end{eqnarray}
After cancelations we arrive at the condition (\ref{cond2_app}).

If one uses the approximation (\ref{Fapp}) for the CGF,
the solution of the saddle point equation (\ref{saddle_app}) becomes
\begin{eqnarray}
\chi_0 = \frac{1}{i\alpha}\ln\left(\frac{N}{2\alpha\Gamma_+ t} + \sqrt{ \frac{\Gamma_-}{\Gamma_+}  +  \frac{N^2}{4\alpha^2\Gamma_+^2 t^2}}\right).
\label{chi0app}
\end{eqnarray}
Hence
\begin{eqnarray}
&& {\cal F}_0(\chi_{0})t-i\chi_{0} N = \sqrt{ 4\Gamma_+\Gamma_- t^2  +  \frac{N^2}{\alpha^2 }}
\nonumber\\ &&
-\, \frac{N}{\alpha}\ln\left( \frac{N}{2\alpha\sqrt{\Gamma_+\Gamma_-} t} + \sqrt{ 1  +  \frac{N^2}{4\alpha^2\Gamma_+\Gamma_- t^2}} \right)
\nonumber\\ &&
-\, (\Gamma_+ + \Gamma_-)t   + \frac{N}{2\alpha}\ln\frac{\Gamma_+}{\Gamma_-}.
\end{eqnarray}
This function has the maximum at time $t=N/{\cal C}_1$, where $\chi_0=0$. Perfoming Taylor expansion around this point,
we find
\begin{eqnarray}
\chi_0 = \frac{\sqrt{{\cal C}_1{\cal C}_3}}{{\cal C}_2}\left( \frac{N}{{\cal C}_1t} -1\right) + {\cal O}\bigg[\left( \frac{N}{{\cal C}_1t} -1\right)^3\bigg],
\label{Taylor_chi_0}
\end{eqnarray}
\begin{eqnarray}
&& {\cal F}_0(\chi_0)t-i\chi_0 N = \frac{N}{2}\sqrt{\frac{{\cal C}_1}{{\cal C}_3}}\ln\frac{{\cal C}_2+\sqrt{{\cal C}_1{\cal C}_3}}{{\cal C}_2-\sqrt{{\cal C}_1{\cal C}_3}}
\nonumber\\ &&
-\, \frac{{\cal C}_1^2 t}{2{\cal C}_2}\left( \frac{N}{{\cal C}_1t} -1\right)^2
+ {\cal O}\bigg[\left( \frac{N}{{\cal C}_1t} -1\right)^3\bigg].
\label{Taylor_F_0}
\end{eqnarray}
In the long time limit, $t\gg N/{\cal C}_1$, we can use Taylor expansion in $\chi$. In this case, the
the difference between the exact CGF (\ref{FCS}) and the approximate one (\ref{Fapp})
appears only in the fourth order of the Taylor expansion in $\chi$, and hence
\begin{eqnarray}
\delta{\cal F}(\chi_0)
= \frac{\chi_0^4}{24} \frac{{\cal C}_1{\cal C}_4 - {\cal C}_2{\cal C}_3}{{\cal C}_1} .
\end{eqnarray}
With the aid of the expansions (\ref{Taylor_chi_0},~\ref{Taylor_F_0}) we write the condition (\ref{cond3}) in the form 
given in the main text
\begin{eqnarray}
\frac{{\cal C}_1|{\cal C}_1{\cal C}_4 - {\cal C}_2{\cal C}_3|}{12{\cal C}_2^3} \left( \frac{N}{{\cal C}_1t} -1\right)^2 \lesssim 1.
\label{cond3_app}
\end{eqnarray}

\begin{figure*}[!ht]
	\includegraphics[width= 1.5 \columnwidth]{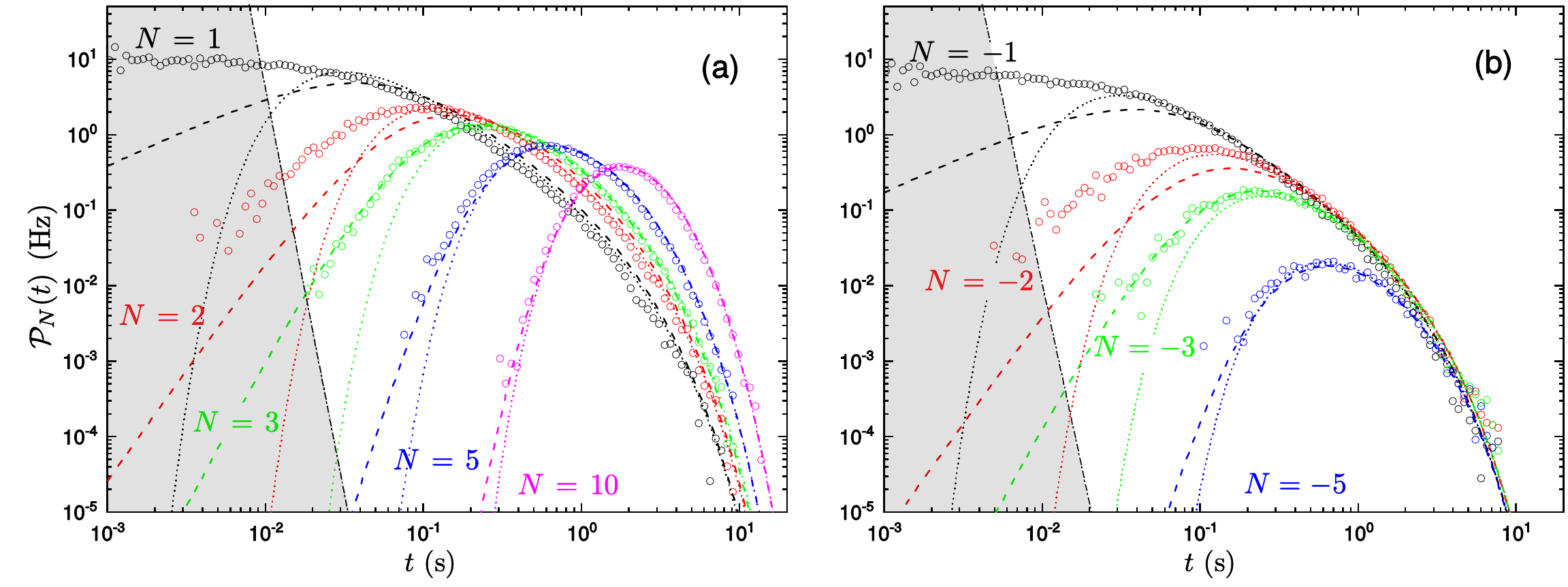}
	\caption{\textbf{First passage time distribution} at bias voltage $V_{\rm b}=65$ $\mu$V 
         for the thresholds $N$  in the direction of flow (a) and  against the flow (b).
		The symbols are the experimental data, solid lines show  the diffusion approximation (\ref{P_diff}),
		dashed lines are the predictions of the Eq.~\eqref{PC3_app}, and dotted black lines mark the validity conditions~\eqref{cond3_app} and (\ref{cond2_app}),
        which are very close for this bias voltage.
        For this bias voltage the rates are 
        $\Gamma_{00}^{01}=120$ Hz, $\Gamma_{01}^{00}=71$ Hz,  
        $\Gamma_{00}^{10}=51$ Hz, $\Gamma_{10}^{00}=30$ Hz,
        $\Gamma_{01}^{11}=34$ Hz, $\Gamma_{11}^{01}=35$ Hz, $\Gamma_{10}^{11}=88$ Hz, $\Gamma_{11}^{10}=98$ Hz, $\Gamma_{01}^{10}=46$ Hz, $\Gamma_{10}^{01}=21$ Hz.
        The corresponding values of the cumulants are ${\cal C}_1=3.72$ Hz, ${\cal C}_1=10.42$ Hz, ${\cal C}_3=1.17$ Hz, and ${\cal C}_4=3.1$ Hz. 
        }
	\label{fig:fptd_comp2}
\end{figure*}

\section{Comparison of the approximate theoretical models at different bias voltages}
\label{comp_res}

It is interesting to compare the diffusion approximation (\ref{P_diff}) with the more accurate one  (\ref{PC3_app}). 
This comparison is made in Fig. \ref{fig:fptd_comp} for the bias voltage  $V_{\rm b}=90$ $\mu$V (the same value as in the main text),
and in Fig. \ref{fig:fptd_comp2} --- for $V_{\rm b}=65$ $\mu$V.
We observe that both approximations work well at long times since both validity conditions
(\ref{cond2_app}) and (\ref{cond3_app}) are satisfied in this limit.
However, at shorter times the approximation (\ref{PC3_app}) systematically fits the data better, as expected.
Indeed, at $V_{\rm b}=90$ $\mu$V, for example, the validity condition (\ref{cond2_app}) of the diffusion approximation at short times may be written in the form
$
t\gtrsim 53\times |N|\;{\rm ms}\; {\rm for}\; N\leq -1;\;\;\; t\gtrsim 36\times N\;{\rm ms}\; {\rm for}\; N\geq 1.
$
At the same time, the approximation (\ref{PC3_app}) should be valid provided
$
t\gtrsim 6.9\times |N|\;{\rm ms}\; {\rm for}\; N\leq -1;\;\;\; t\gtrsim 6.5\times N\;{\rm ms}\; {\rm for}\; N\geq 1.
$
These boundaries are indicated by black dashed-dotted lines in Figs. \ref{fig:fptd_comp}~(a,b).
The same analysis for $V_{\rm b}=65$ $\mu$V reveals that the conditions (\ref{cond2_app}) and (\ref{PC3_app})
give very similar boundaries, which are shown by dashed-dotted lines in Figs. \ref{fig:fptd_comp2}~(a,b).
However, also in this case the approximation (\ref{PC3_app}) works better.

At $V_{\rm b}=90$ $\mu$V we find $e^*=e\sqrt{{\cal C}_1/{\cal C}_3}=1.45 e$, and $|N^*|=|N|$ for $N=\pm 1$.
That is why at this bias the approximation (\ref{PC3_app}) works well even for small values of the threshold,
see Fig. \ref{fig:fptd_comp}.
In contrast, at $V_{\rm b}=65$ $\mu$V we find $e^*=1.78 e$ and $N^* = 2$ for $N=1$. 
Therefore the approximation (\ref{PC3_app}) becomes applicable only for $|N|\geq 3$, where the short time
part of the distribution becomes experimentally inaccessible due to its low statistical weight.


\begin{thebibliography}{53}%
	\makeatletter
	\providecommand \@ifxundefined [1]{%
		\@ifx{#1\undefined}
	}%
	\providecommand \@ifnum [1]{%
		\ifnum #1\expandafter \@firstoftwo
		\else \expandafter \@secondoftwo
		\fi
	}%
	\providecommand \@ifx [1]{%
		\ifx #1\expandafter \@firstoftwo
		\else \expandafter \@secondoftwo
		\fi
	}%
	\providecommand \natexlab [1]{#1}%
	\providecommand \enquote  [1]{``#1''}%
	\providecommand \bibnamefont  [1]{#1}%
	\providecommand \bibfnamefont [1]{#1}%
	\providecommand \citenamefont [1]{#1}%
	\providecommand \href@noop [0]{\@secondoftwo}%
	\providecommand \href [0]{\begingroup \@sanitize@url \@href}%
	\providecommand \@href[1]{\@@startlink{#1}\@@href}%
	\providecommand \@@href[1]{\endgroup#1\@@endlink}%
	\providecommand \@sanitize@url [0]{\catcode `\\12\catcode `\$12\catcode
		`\&12\catcode `\#12\catcode `\^12\catcode `\_12\catcode `\%12\relax}%
	\providecommand \@@startlink[1]{}%
	\providecommand \@@endlink[0]{}%
	\providecommand \url  [0]{\begingroup\@sanitize@url \@url }%
	\providecommand \@url [1]{\endgroup\@href {#1}{\urlprefix }}%
	\providecommand \urlprefix  [0]{URL }%
	\providecommand \Eprint [0]{\href }%
	\providecommand \doibase [0]{http://dx.doi.org/}%
	\providecommand \selectlanguage [0]{\@gobble}%
	\providecommand \bibinfo  [0]{\@secondoftwo}%
	\providecommand \bibfield  [0]{\@secondoftwo}%
	\providecommand \translation [1]{[#1]}%
	\providecommand \BibitemOpen [0]{}%
	\providecommand \bibitemStop [0]{}%
	\providecommand \bibitemNoStop [0]{.\EOS\space}%
	\providecommand \EOS [0]{\spacefactor3000\relax}%
	\providecommand \BibitemShut  [1]{\csname bibitem#1\endcsname}%
	\let\auto@bib@innerbib\@empty
	\bibitem [{\citenamefont {Smoluchowski}(1915)}]{Smoluchowski}%
	\BibitemOpen
	\bibfield  {author} {\bibinfo {author} {\bibfnamefont {M.~V.}\ \bibnamefont
			{Smoluchowski}},\ }\href@noop {} {\bibfield  {journal} {\bibinfo  {journal}
			{Physikalische Zeitschrift}\ }\textbf {\bibinfo {volume} {16}},\ \bibinfo
		{pages} {318} (\bibinfo {year} {1915})}\BibitemShut {NoStop}%
	\bibitem [{\citenamefont {Schrodinger}(1915)}]{Schrodinger}%
	\BibitemOpen
	\bibfield  {author} {\bibinfo {author} {\bibfnamefont {E.}~\bibnamefont
			{Schrodinger}},\ }\href@noop {} {\bibfield  {journal} {\bibinfo  {journal}
			{Physikalische Zeitschrift}\ }\textbf {\bibinfo {volume} {16}},\ \bibinfo
		{pages} {289} (\bibinfo {year} {1915})}\BibitemShut {NoStop}%
	\bibitem [{\citenamefont {Redner}(2001)}]{Redner}%
	\BibitemOpen
	\bibfield  {author} {\bibinfo {author} {\bibfnamefont {S.}~\bibnamefont
			{Redner}},\ }\href
	{https://www.cambridge.org/fi/academic/subjects/physics/statistical-physics/guide-first-passage-processes?format=HB&isbn=9780521652483}
	{\emph {\bibinfo {title} {{A Guide to First-Passage Processes}}}}\ (\bibinfo
	{publisher} {Cambridge University Press},\ \bibinfo {address} {Cambridge,
		UK},\ \bibinfo {year} {2001})\BibitemShut {NoStop}%
	\bibitem [{\citenamefont {Metzler}\ \emph {et~al.}(2014)\citenamefont
		{Metzler}, \citenamefont {Redner},\ and\ \citenamefont {Oshanin}}]{Metzler}%
	\BibitemOpen
	\bibfield  {author} {\bibinfo {author} {\bibfnamefont {R.}~\bibnamefont
			{Metzler}}, \bibinfo {author} {\bibfnamefont {S.}~\bibnamefont {Redner}}, \
		and\ \bibinfo {author} {\bibfnamefont {G.}~\bibnamefont {Oshanin}},\ }\href
	{https://doi.org/10.1142/9104} {\emph {\bibinfo {title} {{First-passage
					phenomena and their applications}}}}\ (\bibinfo  {publisher} {World
		Scientific},\ \bibinfo {year} {2014})\BibitemShut {NoStop}%
	\bibitem [{\citenamefont {Szabo}\ \emph {et~al.}(1980)\citenamefont {Szabo},
		\citenamefont {Schulten},\ and\ \citenamefont {Schulten}}]{Szabo}%
	\BibitemOpen
	\bibfield  {author} {\bibinfo {author} {\bibfnamefont {A.}~\bibnamefont
			{Szabo}}, \bibinfo {author} {\bibfnamefont {K.}~\bibnamefont {Schulten}}, \
		and\ \bibinfo {author} {\bibfnamefont {Z.}~\bibnamefont {Schulten}},\
	}\href@noop {} {\bibfield  {journal} {\bibinfo  {journal} {J. Chem. Phys.}\
	}\textbf {\bibinfo {volume} {72}},\ \bibinfo {pages} {4350} (\bibinfo {year}
	{1980})}\BibitemShut {NoStop}%
\bibitem [{\citenamefont {Zwanzig}\ \emph {et~al.}(1992)\citenamefont
	{Zwanzig}, \citenamefont {Szabo},\ and\ \citenamefont {Bagchi}}]{Zwanzig}%
\BibitemOpen
\bibfield  {author} {\bibinfo {author} {\bibfnamefont {R.}~\bibnamefont
		{Zwanzig}}, \bibinfo {author} {\bibfnamefont {A.}~\bibnamefont {Szabo}}, \
	and\ \bibinfo {author} {\bibfnamefont {B.}~\bibnamefont {Bagchi}},\ }\href
{http://www.pnas.org/content/89/1/20} {\bibfield  {journal} {\bibinfo
		{journal} {Proc. Natl. Acad. Sci. USA}\ }\textbf {\bibinfo {volume} {89}},\
	\bibinfo {pages} {20} (\bibinfo {year} {1992})}\BibitemShut {NoStop}%
\bibitem [{\citenamefont {Galburt}\ \emph {et~al.}(2007)\citenamefont
	{Galburt}, \citenamefont {Grill}, \citenamefont {Wiedmann}, \citenamefont
	{Lubkowska}, \citenamefont {Choy}, \citenamefont {Nogales}, \citenamefont
	{Kashlev},\ and\ \citenamefont {Bustamante}}]{Galburt}%
\BibitemOpen
\bibfield  {author} {\bibinfo {author} {\bibfnamefont {E.~A.}\ \bibnamefont
		{Galburt}}, \bibinfo {author} {\bibfnamefont {S.~W.}\ \bibnamefont {Grill}},
	\bibinfo {author} {\bibfnamefont {A.}~\bibnamefont {Wiedmann}}, \bibinfo
	{author} {\bibfnamefont {L.}~\bibnamefont {Lubkowska}}, \bibinfo {author}
	{\bibfnamefont {J.}~\bibnamefont {Choy}}, \bibinfo {author} {\bibfnamefont
		{E.}~\bibnamefont {Nogales}}, \bibinfo {author} {\bibfnamefont
		{M.}~\bibnamefont {Kashlev}}, \ and\ \bibinfo {author} {\bibfnamefont
		{C.}~\bibnamefont {Bustamante}},\ }\href {\doibase 10.1038/nature05701}
{\bibfield  {journal} {\bibinfo  {journal} {Nature}\ }\textbf {\bibinfo
		{volume} {446}},\ \bibinfo {pages} {820} (\bibinfo {year}
	{2007})}\BibitemShut {NoStop}%
\bibitem [{\citenamefont {Qian}\ and\ \citenamefont {Xie}(2006)}]{Xie}%
\BibitemOpen
\bibfield  {author} {\bibinfo {author} {\bibfnamefont {H.}~\bibnamefont
		{Qian}}\ and\ \bibinfo {author} {\bibfnamefont {X.~S.}\ \bibnamefont {Xie}},\
}\href {\doibase 10.1103/PhysRevE.74.010902} {\bibfield  {journal} {\bibinfo
	{journal} {Phys. Rev. E}\ }\textbf {\bibinfo {volume} {74}},\ \bibinfo
{pages} {010902} (\bibinfo {year} {2006})}\BibitemShut {NoStop}%
\bibitem [{\citenamefont {Bauer}\ and\ \citenamefont {Cornu}(2014)}]{Bauer}%
\BibitemOpen
\bibfield  {author} {\bibinfo {author} {\bibfnamefont {M.}~\bibnamefont
		{Bauer}}\ and\ \bibinfo {author} {\bibfnamefont {F.}~\bibnamefont {Cornu}},\
}\href {https://link.springer.com/article/10.1007/s10955-014-0969-z}
{\bibfield  {journal} {\bibinfo  {journal} {J. Stat. Phys.}\ }\textbf
	{\bibinfo {volume} {155}},\ \bibinfo {pages} {703} (\bibinfo {year}
	{2014})}\BibitemShut {NoStop}%
\bibitem [{\citenamefont {Benichou}\ \emph {et~al.}(2010)\citenamefont
	{Benichou}, \citenamefont {Chevalier}, \citenamefont {Klafter}, \citenamefont
	{Meyer},\ and\ \citenamefont {Voituriez}}]{Benichou}%
\BibitemOpen
\bibfield  {author} {\bibinfo {author} {\bibfnamefont {O.}~\bibnamefont
		{Benichou}}, \bibinfo {author} {\bibfnamefont {C.}~\bibnamefont {Chevalier}},
	\bibinfo {author} {\bibfnamefont {J.}~\bibnamefont {Klafter}}, \bibinfo
	{author} {\bibfnamefont {B.}~\bibnamefont {Meyer}}, \ and\ \bibinfo {author}
	{\bibfnamefont {R.}~\bibnamefont {Voituriez}},\ }\href
{http://dx.doi.org/10.1038/nchem.622} {\bibfield  {journal} {\bibinfo
		{journal} {Nat. Chem.}\ }\textbf {\bibinfo {volume} {2}},\ \bibinfo {pages}
	{472} (\bibinfo {year} {2010})}\BibitemShut {NoStop}%
\bibitem [{\citenamefont {Chandrasekhar}(1943)}]{Ch}%
\BibitemOpen
\bibfield  {author} {\bibinfo {author} {\bibfnamefont {S.}~\bibnamefont
		{Chandrasekhar}},\ }\href {\doibase 10.1086/144518} {\bibfield  {journal}
	{\bibinfo  {journal} {Astrophys J.}\ }\textbf {\bibinfo {volume} {97}},\
	\bibinfo {pages} {263} (\bibinfo {year} {1943})}\BibitemShut {NoStop}%
\bibitem [{\citenamefont {Majumdar}(2005)}]{Satya}%
\BibitemOpen
\bibfield  {author} {\bibinfo {author} {\bibfnamefont {S.~N.}\ \bibnamefont
		{Majumdar}},\ }\href@noop {} {\bibfield  {journal} {\bibinfo  {journal}
		{Curr. Sci.}\ }\textbf {\bibinfo {volume} {89}},\ \bibinfo {pages} {2076}
	(\bibinfo {year} {2005})}\BibitemShut {NoStop}%
\bibitem [{\citenamefont {Wald}(1973)}]{Wald}%
\BibitemOpen
\bibfield  {author} {\bibinfo {author} {\bibfnamefont {A.}~\bibnamefont
		{Wald}},\ }\href@noop {} {\emph {\bibinfo {title} {Sequential analysis}}}\
(\bibinfo  {publisher} {Courier Corporation},\ \bibinfo {year}
{1973})\BibitemShut {NoStop}%
\bibitem [{\citenamefont {Siggia}\ and\ \citenamefont
	{Vergassola}(2013)}]{Vergassola}%
\BibitemOpen
\bibfield  {author} {\bibinfo {author} {\bibfnamefont {E.~D.}\ \bibnamefont
		{Siggia}}\ and\ \bibinfo {author} {\bibfnamefont {M.}~\bibnamefont
		{Vergassola}},\ }\href {\doibase 10.1073/pnas.1314081110} {\bibfield
	{journal} {\bibinfo  {journal} {Proc. Natl. Acad. Sci.}\ }\textbf {\bibinfo
		{volume} {110}},\ \bibinfo {pages} {E3704} (\bibinfo {year}
	{2013})}\BibitemShut {NoStop}%
\bibitem [{\citenamefont {Rold\'an}\ \emph {et~al.}(2015)\citenamefont
	{Rold\'an}, \citenamefont {Neri}, \citenamefont {D\"orpinghaus},
	\citenamefont {Meyr},\ and\ \citenamefont {J\"ulicher}}]{Roldan}%
\BibitemOpen
\bibfield  {author} {\bibinfo {author} {\bibfnamefont {{\'E}.}~\bibnamefont
		{Rold\'an}}, \bibinfo {author} {\bibfnamefont {I.}~\bibnamefont {Neri}},
	\bibinfo {author} {\bibfnamefont {M.}~\bibnamefont {D\"orpinghaus}}, \bibinfo
	{author} {\bibfnamefont {H.}~\bibnamefont {Meyr}}, \ and\ \bibinfo {author}
	{\bibfnamefont {F.}~\bibnamefont {J\"ulicher}},\ }\href {\doibase
	10.1103/PhysRevLett.115.250602} {\bibfield  {journal} {\bibinfo  {journal}
		{Phys. Rev. Lett.}\ }\textbf {\bibinfo {volume} {115}},\ \bibinfo {pages}
	{250602} (\bibinfo {year} {2015})}\BibitemShut {NoStop}%
\bibitem [{\citenamefont {Tejedor}\ \emph {et~al.}(2012)\citenamefont
	{Tejedor}, \citenamefont {Voituriez},\ and\ \citenamefont
	{B\'enichou}}]{Tejedor}%
\BibitemOpen
\bibfield  {author} {\bibinfo {author} {\bibfnamefont {V.}~\bibnamefont
		{Tejedor}}, \bibinfo {author} {\bibfnamefont {R.}~\bibnamefont {Voituriez}},
	\ and\ \bibinfo {author} {\bibfnamefont {O.}~\bibnamefont {B\'enichou}},\
}\href {\doibase 10.1103/PhysRevLett.108.088103} {\bibfield  {journal}
{\bibinfo  {journal} {Phys. Rev. Lett.}\ }\textbf {\bibinfo {volume} {108}},\
\bibinfo {pages} {088103} (\bibinfo {year} {2012})}\BibitemShut {NoStop}%
\bibitem [{\citenamefont {Mejía-Monasterio}\ \emph {et~al.}(2011)\citenamefont
	{Mejía-Monasterio}, \citenamefont {Oshanin},\ and\ \citenamefont
	{Schehr}}]{Carlos}%
\BibitemOpen
\bibfield  {author} {\bibinfo {author} {\bibfnamefont {C.}~\bibnamefont
		{Mejía-Monasterio}}, \bibinfo {author} {\bibfnamefont {G.}~\bibnamefont
		{Oshanin}}, \ and\ \bibinfo {author} {\bibfnamefont {G.}~\bibnamefont
		{Schehr}},\ }\href {http://dx.doi.org/10.1063/1.4990866} {\bibfield
	{journal} {\bibinfo  {journal} {J. Stat. Phys.}\ }\textbf {\bibinfo {volume}
		{2011}},\ \bibinfo {pages} {P06022} (\bibinfo {year} {2011})}\BibitemShut
{NoStop}%
\bibitem [{\citenamefont {R.~Chicheportiche}(2014)}]{Bouchaud}%
\BibitemOpen
\bibfield  {author} {\bibinfo {author} {\bibfnamefont {J.-P.~B.}\
		\bibnamefont {R.~Chicheportiche}},\ }\href@noop {} {\emph {\bibinfo {title}
		{First-passage Phenomena and Their Applications}}}\ (\bibinfo {year}
{2014})\BibitemShut {NoStop}%
\bibitem [{\citenamefont {Perell\'o}\ \emph {et~al.}(2011)\citenamefont
	{Perell\'o}, \citenamefont {Guti\'errez-Roig},\ and\ \citenamefont
	{Masoliver}}]{Perello}%
\BibitemOpen
\bibfield  {author} {\bibinfo {author} {\bibfnamefont {J.}~\bibnamefont
		{Perell\'o}}, \bibinfo {author} {\bibfnamefont {M.}~\bibnamefont
		{Guti\'errez-Roig}}, \ and\ \bibinfo {author} {\bibfnamefont
		{J.}~\bibnamefont {Masoliver}},\ }\href {\doibase 10.1103/PhysRevE.84.066110}
{\bibfield  {journal} {\bibinfo  {journal} {Phys. Rev. E}\ }\textbf {\bibinfo
		{volume} {84}},\ \bibinfo {pages} {066110} (\bibinfo {year}
	{2011})}\BibitemShut {NoStop}%
\bibitem [{\citenamefont {Saito}\ and\ \citenamefont {Dhar}(2016)}]{SD}%
\BibitemOpen
\bibfield  {author} {\bibinfo {author} {\bibfnamefont {K.}~\bibnamefont
		{Saito}}\ and\ \bibinfo {author} {\bibfnamefont {A.}~\bibnamefont {Dhar}},\
}\href {http://stacks.iop.org/0295-5075/114/i=5/a=50004} {\bibfield
{journal} {\bibinfo  {journal} {Europhys. Lett.}\ }\textbf {\bibinfo {volume}
	{114}},\ \bibinfo {pages} {50004} (\bibinfo {year} {2016})}\BibitemShut
{NoStop}%
\bibitem [{\citenamefont {Ptaszy\ifmmode~\acute{n}\else
		\'{n}\fi{}ski}(2018)}]{Ptaszynski}%
\BibitemOpen
\bibfield  {author} {\bibinfo {author} {\bibfnamefont {K.}~\bibnamefont
		{Ptaszy\ifmmode~\acute{n}\else \'{n}\fi{}ski}},\ }\href {\doibase
	10.1103/PhysRevE.97.012127} {\bibfield  {journal} {\bibinfo  {journal} {Phys.
			Rev. E}\ }\textbf {\bibinfo {volume} {97}},\ \bibinfo {pages} {012127}
	(\bibinfo {year} {2018})}\BibitemShut {NoStop}%
\bibitem [{\citenamefont {Neri}\ \emph {et~al.}(2017)\citenamefont {Neri},
	\citenamefont {Rold\'an},\ and\ \citenamefont {J\"ulicher}}]{PRX}%
\BibitemOpen
\bibfield  {author} {\bibinfo {author} {\bibfnamefont {I.}~\bibnamefont
		{Neri}}, \bibinfo {author} {\bibfnamefont {{\'E}.}~\bibnamefont {Rold\'an}},
	\ and\ \bibinfo {author} {\bibfnamefont {F.}~\bibnamefont {J\"ulicher}},\
}\href {\doibase 10.1103/PhysRevX.7.011019} {\bibfield  {journal} {\bibinfo
	{journal} {Phys. Rev. X}\ }\textbf {\bibinfo {volume} {7}},\ \bibinfo {pages}
{011019} (\bibinfo {year} {2017})}\BibitemShut {NoStop}%
\bibitem [{\citenamefont {Gingrich}\ and\ \citenamefont
	{Horowitz}(2017)}]{Jordan}%
\BibitemOpen
\bibfield  {author} {\bibinfo {author} {\bibfnamefont {T.~R.}\ \bibnamefont
		{Gingrich}}\ and\ \bibinfo {author} {\bibfnamefont {J.~M.}\ \bibnamefont
		{Horowitz}},\ }\href {\doibase 10.1103/PhysRevLett.119.170601} {\bibfield
	{journal} {\bibinfo  {journal} {Phys. Rev. Lett.}\ }\textbf {\bibinfo
		{volume} {119}},\ \bibinfo {pages} {170601} (\bibinfo {year}
	{2017})}\BibitemShut {NoStop}%
\bibitem [{\citenamefont {Garrahan}(2017)}]{Garrahan}%
\BibitemOpen
\bibfield  {author} {\bibinfo {author} {\bibfnamefont {J.~P.}\ \bibnamefont
		{Garrahan}},\ }\href {\doibase 10.1103/PhysRevE.95.032134} {\bibfield
	{journal} {\bibinfo  {journal} {Phys. Rev. E}\ }\textbf {\bibinfo {volume}
		{95}},\ \bibinfo {pages} {032134} (\bibinfo {year} {2017})}\BibitemShut
{NoStop}%
\bibitem [{\citenamefont {H{\"a}nggi}\ \emph {et~al.}(1990)\citenamefont
	{H{\"a}nggi}, \citenamefont {Talkner},\ and\ \citenamefont
	{Borkovec}}]{Hanggi}%
\BibitemOpen
\bibfield  {author} {\bibinfo {author} {\bibfnamefont {P.}~\bibnamefont
		{H{\"a}nggi}}, \bibinfo {author} {\bibfnamefont {P.}~\bibnamefont {Talkner}},
	\ and\ \bibinfo {author} {\bibfnamefont {M.}~\bibnamefont {Borkovec}},\
}\href {\doibase 10.1103/RevModPhys.62.251} {\bibfield  {journal} {\bibinfo
	{journal} {Rev. Mod. Phys.}\ }\textbf {\bibinfo {volume} {62}},\ \bibinfo
{pages} {251} (\bibinfo {year} {1990})}\BibitemShut {NoStop}%
\bibitem [{\citenamefont {Albert}\ \emph {et~al.}(2011)\citenamefont {Albert},
	\citenamefont {Flindt},\ and\ \citenamefont {B\"uttiker}}]{Flindt1}%
\BibitemOpen
\bibfield  {author} {\bibinfo {author} {\bibfnamefont {M.}~\bibnamefont
		{Albert}}, \bibinfo {author} {\bibfnamefont {C.}~\bibnamefont {Flindt}}, \
	and\ \bibinfo {author} {\bibfnamefont {M.}~\bibnamefont {B\"uttiker}},\
}\href {\doibase 10.1103/PhysRevLett.107.086805} {\bibfield  {journal}
{\bibinfo  {journal} {Phys. Rev. Lett.}\ }\textbf {\bibinfo {volume} {107}},\
\bibinfo {pages} {086805} (\bibinfo {year} {2011})}\BibitemShut {NoStop}%
\bibitem [{\citenamefont {Dasenbrook}\ \emph {et~al.}(2014)\citenamefont
	{Dasenbrook}, \citenamefont {Flindt},\ and\ \citenamefont
	{B\"uttiker}}]{Flindt2}%
\BibitemOpen
\bibfield  {author} {\bibinfo {author} {\bibfnamefont {D.}~\bibnamefont
		{Dasenbrook}}, \bibinfo {author} {\bibfnamefont {C.}~\bibnamefont {Flindt}},
	\ and\ \bibinfo {author} {\bibfnamefont {M.}~\bibnamefont {B\"uttiker}},\
}\href {\doibase 10.1103/PhysRevLett.112.146801} {\bibfield  {journal}
{\bibinfo  {journal} {Phys. Rev. Lett.}\ }\textbf {\bibinfo {volume} {112}},\
\bibinfo {pages} {146801} (\bibinfo {year} {2014})}\BibitemShut {NoStop}%
\bibitem [{\citenamefont {Lu}\ \emph {et~al.}(2003)\citenamefont {Lu},
	\citenamefont {Ji}, \citenamefont {Pfeiffer}, \citenamefont {West},\ and\
	\citenamefont {Rimberg}}]{Lu}%
\BibitemOpen
\bibfield  {author} {\bibinfo {author} {\bibfnamefont {W.}~\bibnamefont
		{Lu}}, \bibinfo {author} {\bibfnamefont {Z.}~\bibnamefont {Ji}}, \bibinfo
	{author} {\bibfnamefont {L.}~\bibnamefont {Pfeiffer}}, \bibinfo {author}
	{\bibfnamefont {K.~W.}\ \bibnamefont {West}}, \ and\ \bibinfo {author}
	{\bibfnamefont {A.~J.}\ \bibnamefont {Rimberg}},\ }\href
{http://dx.doi.org/10.1038/nature01642} {\bibfield  {journal} {\bibinfo
		{journal} {Nature}\ }\textbf {\bibinfo {volume} {423}},\ \bibinfo {pages}
	{422 EP } (\bibinfo {year} {2003})}\BibitemShut {NoStop}%
\bibitem [{\citenamefont {Fujisawa}\ \emph {et~al.}(2006)\citenamefont
	{Fujisawa}, \citenamefont {Hayashi}, \citenamefont {Tomita},\ and\
	\citenamefont {Hirayama}}]{Fujisawa}%
\BibitemOpen
\bibfield  {author} {\bibinfo {author} {\bibfnamefont {T.}~\bibnamefont
		{Fujisawa}}, \bibinfo {author} {\bibfnamefont {T.}~\bibnamefont {Hayashi}},
	\bibinfo {author} {\bibfnamefont {R.}~\bibnamefont {Tomita}}, \ and\ \bibinfo
	{author} {\bibfnamefont {Y.}~\bibnamefont {Hirayama}},\ }\href {\doibase
	10.1126/science.1126788} {\bibfield  {journal} {\bibinfo  {journal}
		{Science}\ }\textbf {\bibinfo {volume} {312}},\ \bibinfo {pages} {1634}
	(\bibinfo {year} {2006})}\BibitemShut {NoStop}%
\bibitem [{\citenamefont {Gustavsson}\ \emph {et~al.}(2006)\citenamefont
	{Gustavsson}, \citenamefont {Leturcq}, \citenamefont
	{Simovi\ifmmode~\check{c}\else \v{c}\fi{}}, \citenamefont {Schleser},
	\citenamefont {Ihn}, \citenamefont {Studerus}, \citenamefont {Ensslin},
	\citenamefont {Driscoll},\ and\ \citenamefont {Gossard}}]{Gustavsson}%
\BibitemOpen
\bibfield  {author} {\bibinfo {author} {\bibfnamefont {S.}~\bibnamefont
		{Gustavsson}}, \bibinfo {author} {\bibfnamefont {R.}~\bibnamefont {Leturcq}},
	\bibinfo {author} {\bibfnamefont {B.}~\bibnamefont
		{Simovi\ifmmode~\check{c}\else \v{c}\fi{}}}, \bibinfo {author} {\bibfnamefont
		{R.}~\bibnamefont {Schleser}}, \bibinfo {author} {\bibfnamefont
		{T.}~\bibnamefont {Ihn}}, \bibinfo {author} {\bibfnamefont {P.}~\bibnamefont
		{Studerus}}, \bibinfo {author} {\bibfnamefont {K.}~\bibnamefont {Ensslin}},
	\bibinfo {author} {\bibfnamefont {D.~C.}\ \bibnamefont {Driscoll}}, \ and\
	\bibinfo {author} {\bibfnamefont {A.~C.}\ \bibnamefont {Gossard}},\ }\href
{\doibase 10.1103/PhysRevLett.96.076605} {\bibfield  {journal} {\bibinfo
		{journal} {Phys. Rev. Lett.}\ }\textbf {\bibinfo {volume} {96}},\ \bibinfo
	{pages} {076605} (\bibinfo {year} {2006})}\BibitemShut {NoStop}%
\bibitem [{\citenamefont {Flindt}\ \emph {et~al.}(2009)\citenamefont {Flindt},
	\citenamefont {Fricke}, \citenamefont {Hohls}, \citenamefont {Novotn{\'y}},
	\citenamefont {Neto{\v c}n{\'y}}, \citenamefont {Brandes},\ and\
	\citenamefont {Haug}}]{Haug}%
\BibitemOpen
\bibfield  {author} {\bibinfo {author} {\bibfnamefont {C.}~\bibnamefont
		{Flindt}}, \bibinfo {author} {\bibfnamefont {C.}~\bibnamefont {Fricke}},
	\bibinfo {author} {\bibfnamefont {F.}~\bibnamefont {Hohls}}, \bibinfo
	{author} {\bibfnamefont {T.}~\bibnamefont {Novotn{\'y}}}, \bibinfo {author}
	{\bibfnamefont {K.}~\bibnamefont {Neto{\v c}n{\'y}}}, \bibinfo {author}
	{\bibfnamefont {T.}~\bibnamefont {Brandes}}, \ and\ \bibinfo {author}
	{\bibfnamefont {R.~J.}\ \bibnamefont {Haug}},\ }\href {\doibase
	10.1073/pnas.0901002106} {\bibfield  {journal} {\bibinfo  {journal} {Proc.
			Natl. Acad. Sci.}\ }\textbf {\bibinfo {volume} {106}},\ \bibinfo {pages}
	{10116} (\bibinfo {year} {2009})}\BibitemShut {NoStop}%
\bibitem [{\citenamefont {Majumdar}(2010)}]{Majumdar}%
\BibitemOpen
\bibfield  {author} {\bibinfo {author} {\bibfnamefont {S.~N.}\ \bibnamefont
		{Majumdar}},\ }\href {\doibase https://doi.org/10.1016/j.physa.2010.01.021}
{\bibfield  {journal} {\bibinfo  {journal} {Physica A: Statistical Mechanics
			and its Applications}\ }\textbf {\bibinfo {volume} {389}},\ \bibinfo {pages}
	{4299 } (\bibinfo {year} {2010})}\BibitemShut {NoStop}%
\bibitem [{\citenamefont {D{\"{o}}rpinghaus}\ \emph {et~al.}(2017)\citenamefont
	{D{\"{o}}rpinghaus}, \citenamefont {Rold\'an}, \citenamefont {Neri},
	\citenamefont {Meyr},\ and\ \citenamefont {J{\"{u}}licher}}]{Meik1}%
\BibitemOpen
\bibfield  {author} {\bibinfo {author} {\bibfnamefont {M.}~\bibnamefont
		{D{\"{o}}rpinghaus}}, \bibinfo {author} {\bibfnamefont {{\'E}.}~\bibnamefont
		{Rold\'an}}, \bibinfo {author} {\bibfnamefont {I.}~\bibnamefont {Neri}},
	\bibinfo {author} {\bibfnamefont {H.}~\bibnamefont {Meyr}}, \ and\ \bibinfo
	{author} {\bibfnamefont {F.}~\bibnamefont {J{\"{u}}licher}},\ }in\ \href
{\doibase 10.1109/ISIT.2017.8007090} {\emph {\bibinfo {booktitle} {IEEE
			International Symposium on Information Theory (ISIT)}}}\ (\bibinfo {year}
{2017})\BibitemShut {NoStop}%
\bibitem [{\citenamefont {D{\"{o}}rpinghaus}\ \emph {et~al.}(2018)\citenamefont
	{D{\"{o}}rpinghaus}, \citenamefont {Neri}, \citenamefont {Rold\'an},
	\citenamefont {Meyr},\ and\ \citenamefont {J{\"{u}}licher}}]{Meik2}%
\BibitemOpen
\bibfield  {author} {\bibinfo {author} {\bibfnamefont {M.}~\bibnamefont
		{D{\"{o}}rpinghaus}}, \bibinfo {author} {\bibfnamefont {I.}~\bibnamefont
		{Neri}}, \bibinfo {author} {\bibfnamefont {{\'E}.}~\bibnamefont {Rold\'an}},
	\bibinfo {author} {\bibfnamefont {H.}~\bibnamefont {Meyr}}, \ and\ \bibinfo
	{author} {\bibfnamefont {F.}~\bibnamefont {J{\"{u}}licher}},\ }\href
{http://arxiv.org/abs/1801.01574} {\bibfield  {journal} {\bibinfo  {journal}
		{arXiv:1801.01574}\ } (\bibinfo {year} {2018})}\BibitemShut {NoStop}%
\bibitem [{\citenamefont {Singh}\ \emph {et~al.}(2017)\citenamefont {Singh},
	\citenamefont {Rold{\'a}n}, \citenamefont {Neri}, \citenamefont {Khaymovich},
	\citenamefont {Golubev}, \citenamefont {Maisi}, \citenamefont {Peltonen},
	\citenamefont {J{\"u}licher},\ and\ \citenamefont {Pekola}}]{Singh}%
\BibitemOpen
\bibfield  {author} {\bibinfo {author} {\bibfnamefont {S.}~\bibnamefont
		{Singh}}, \bibinfo {author} {\bibfnamefont {{\'E}.}~\bibnamefont
		{Rold{\'a}n}}, \bibinfo {author} {\bibfnamefont {I.}~\bibnamefont {Neri}},
	\bibinfo {author} {\bibfnamefont {I.~M.}\ \bibnamefont {Khaymovich}},
	\bibinfo {author} {\bibfnamefont {D.~S.}\ \bibnamefont {Golubev}}, \bibinfo
	{author} {\bibfnamefont {V.~F.}\ \bibnamefont {Maisi}}, \bibinfo {author}
	{\bibfnamefont {J.~T.}\ \bibnamefont {Peltonen}}, \bibinfo {author}
	{\bibfnamefont {F.}~\bibnamefont {J{\"u}licher}}, \ and\ \bibinfo {author}
	{\bibfnamefont {J.~P.}\ \bibnamefont {Pekola}},\ }\href@noop {} {\bibfield
	{journal} {\bibinfo  {journal} {arXiv:1712.01693}\ } (\bibinfo {year}
	{2017})}\BibitemShut {NoStop}%
\bibitem [{\citenamefont {Tobiska}\ and\ \citenamefont {Nazarov}(2005)}]{TN}%
\BibitemOpen
\bibfield  {author} {\bibinfo {author} {\bibfnamefont {J.}~\bibnamefont
		{Tobiska}}\ and\ \bibinfo {author} {\bibfnamefont {Y.~V.}\ \bibnamefont
		{Nazarov}},\ }\href {\doibase 10.1103/PhysRevB.72.235328} {\bibfield
	{journal} {\bibinfo  {journal} {Phys. Rev. B}\ }\textbf {\bibinfo {volume}
		{72}},\ \bibinfo {pages} {235328} (\bibinfo {year} {2005})}\BibitemShut
{NoStop}%
\bibitem [{\citenamefont {F\"orster}\ and\ \citenamefont
	{B\"uttiker}(2008)}]{Buttiker}%
\BibitemOpen
\bibfield  {author} {\bibinfo {author} {\bibfnamefont {H.}~\bibnamefont
		{F\"orster}}\ and\ \bibinfo {author} {\bibfnamefont {M.}~\bibnamefont
		{B\"uttiker}},\ }\href {\doibase 10.1103/PhysRevLett.101.136805} {\bibfield
	{journal} {\bibinfo  {journal} {Phys. Rev. Lett.}\ }\textbf {\bibinfo
		{volume} {101}},\ \bibinfo {pages} {136805} (\bibinfo {year}
	{2008})}\BibitemShut {NoStop}%
\bibitem [{\citenamefont {Utsumi}\ and\ \citenamefont {Saito}(2009)}]{Utsumi}%
\BibitemOpen
\bibfield  {author} {\bibinfo {author} {\bibfnamefont {Y.}~\bibnamefont
		{Utsumi}}\ and\ \bibinfo {author} {\bibfnamefont {K.}~\bibnamefont {Saito}},\
}\href {\doibase 10.1103/PhysRevB.79.235311} {\bibfield  {journal} {\bibinfo
	{journal} {Phys. Rev. B}\ }\textbf {\bibinfo {volume} {79}},\ \bibinfo
{pages} {235311} (\bibinfo {year} {2009})}\BibitemShut {NoStop}%
\bibitem [{\citenamefont {Andrieux}\ \emph {et~al.}(2009)\citenamefont
	{Andrieux}, \citenamefont {Gaspard}, \citenamefont {Monnai},\ and\
	\citenamefont {Tasaki}}]{Tasaki}%
\BibitemOpen
\bibfield  {author} {\bibinfo {author} {\bibfnamefont {D.}~\bibnamefont
		{Andrieux}}, \bibinfo {author} {\bibfnamefont {P.}~\bibnamefont {Gaspard}},
	\bibinfo {author} {\bibfnamefont {T.}~\bibnamefont {Monnai}}, \ and\ \bibinfo
	{author} {\bibfnamefont {S.}~\bibnamefont {Tasaki}},\ }\href
{http://stacks.iop.org/1367-2630/11/i=4/a=043014} {\bibfield  {journal}
	{\bibinfo  {journal} {New J. Phys.}\ }\textbf {\bibinfo {volume} {11}},\
	\bibinfo {pages} {043014} (\bibinfo {year} {2009})}\BibitemShut {NoStop}%
\bibitem [{\citenamefont {Utsumi}\ \emph {et~al.}(2010)\citenamefont {Utsumi},
	\citenamefont {Golubev}, \citenamefont {Marthaler}, \citenamefont {Saito},
	\citenamefont {Fujisawa},\ and\ \citenamefont {Sch\"on}}]{FT1}%
\BibitemOpen
\bibfield  {author} {\bibinfo {author} {\bibfnamefont {Y.}~\bibnamefont
		{Utsumi}}, \bibinfo {author} {\bibfnamefont {D.~S.}\ \bibnamefont {Golubev}},
	\bibinfo {author} {\bibfnamefont {M.}~\bibnamefont {Marthaler}}, \bibinfo
	{author} {\bibfnamefont {K.}~\bibnamefont {Saito}}, \bibinfo {author}
	{\bibfnamefont {T.}~\bibnamefont {Fujisawa}}, \ and\ \bibinfo {author}
	{\bibfnamefont {G.}~\bibnamefont {Sch\"on}},\ }\href {\doibase
	10.1103/PhysRevB.81.125331} {\bibfield  {journal} {\bibinfo  {journal} {Phys.
			Rev. B}\ }\textbf {\bibinfo {volume} {81}},\ \bibinfo {pages} {125331}
	(\bibinfo {year} {2010})}\BibitemShut {NoStop}%
\bibitem [{\citenamefont {K\"ung}\ \emph {et~al.}(2012)\citenamefont {K\"ung},
	\citenamefont {R\"ossler}, \citenamefont {Beck}, \citenamefont {Marthaler},
	\citenamefont {Golubev}, \citenamefont {Utsumi}, \citenamefont {Ihn},\ and\
	\citenamefont {Ensslin}}]{FT2}%
\BibitemOpen
\bibfield  {author} {\bibinfo {author} {\bibfnamefont {B.}~\bibnamefont
		{K\"ung}}, \bibinfo {author} {\bibfnamefont {C.}~\bibnamefont {R\"ossler}},
	\bibinfo {author} {\bibfnamefont {M.}~\bibnamefont {Beck}}, \bibinfo {author}
	{\bibfnamefont {M.}~\bibnamefont {Marthaler}}, \bibinfo {author}
	{\bibfnamefont {D.~S.}\ \bibnamefont {Golubev}}, \bibinfo {author}
	{\bibfnamefont {Y.}~\bibnamefont {Utsumi}}, \bibinfo {author} {\bibfnamefont
		{T.}~\bibnamefont {Ihn}}, \ and\ \bibinfo {author} {\bibfnamefont
		{K.}~\bibnamefont {Ensslin}},\ }\href {\doibase 10.1103/PhysRevX.2.011001}
{\bibfield  {journal} {\bibinfo  {journal} {Phys. Rev. X}\ }\textbf {\bibinfo
		{volume} {2}},\ \bibinfo {pages} {011001} (\bibinfo {year}
	{2012})}\BibitemShut {NoStop}%
\bibitem [{Note1()}]{Note1}%
\BibitemOpen
\bibinfo {note} {In our analysis, we do not consider cotunneling or Andreev
	tunneling due to high resistance of all junctions.}\BibitemShut {Stop}%
\bibitem [{\citenamefont {Derrida}\ \emph {et~al.}(1992)\citenamefont
	{Derrida}, \citenamefont {Domany},\ and\ \citenamefont {Mukamel}}]{ASEP1}%
\BibitemOpen
\bibfield  {author} {\bibinfo {author} {\bibfnamefont {B.}~\bibnamefont
		{Derrida}}, \bibinfo {author} {\bibfnamefont {E.}~\bibnamefont {Domany}}, \
	and\ \bibinfo {author} {\bibfnamefont {D.}~\bibnamefont {Mukamel}},\ }\href
{\doibase 10.1007/BF01050430} {\bibfield  {journal} {\bibinfo  {journal} {J.
			Stat. Phys.}\ }\textbf {\bibinfo {volume} {69}},\ \bibinfo {pages} {667}
	(\bibinfo {year} {1992})}\BibitemShut {NoStop}%
\bibitem [{\citenamefont {Sch{\"u}tz}\ and\ \citenamefont
	{Domany}(1993)}]{ASEP2}%
\BibitemOpen
\bibfield  {author} {\bibinfo {author} {\bibfnamefont {G.}~\bibnamefont
		{Sch{\"u}tz}}\ and\ \bibinfo {author} {\bibfnamefont {E.}~\bibnamefont
		{Domany}},\ }\href {\doibase 10.1007/BF01048050} {\bibfield  {journal}
	{\bibinfo  {journal} {J. Stat. Phys.}\ }\textbf {\bibinfo {volume} {72}},\
	\bibinfo {pages} {277} (\bibinfo {year} {1993})}\BibitemShut {NoStop}%
\bibitem [{\citenamefont {Golinelli}\ and\ \citenamefont
	{Mallick}(2006)}]{ASEP3}%
\BibitemOpen
\bibfield  {author} {\bibinfo {author} {\bibfnamefont {O.}~\bibnamefont
		{Golinelli}}\ and\ \bibinfo {author} {\bibfnamefont {K.}~\bibnamefont
		{Mallick}},\ }\href {http://stacks.iop.org/0305-4470/39/i=41/a=S03}
{\bibfield  {journal} {\bibinfo  {journal} {J. Phys. A: Math. Gen.}\ }\textbf
	{\bibinfo {volume} {39}},\ \bibinfo {pages} {12679} (\bibinfo {year}
	{2006})}\BibitemShut {NoStop}%
\bibitem [{\citenamefont {Derrida}(2007)}]{ASEP4}%
\BibitemOpen
\bibfield  {author} {\bibinfo {author} {\bibfnamefont {B.}~\bibnamefont
		{Derrida}},\ }\href {http://stacks.iop.org/1742-5468/2007/i=07/a=P07023}
{\bibfield  {journal} {\bibinfo  {journal} {J. Stat. Mech.}\ }\textbf
	{\bibinfo {volume} {2007}},\ \bibinfo {pages} {P07023} (\bibinfo {year}
	{2007})}\BibitemShut {NoStop}%
\bibitem [{\citenamefont {Cates}\ and\ \citenamefont {Evans}(2000)}]{ASEP5}%
\BibitemOpen
\bibfield  {author} {\bibinfo {author} {\bibfnamefont {M.~E.}\ \bibnamefont
		{Cates}}\ and\ \bibinfo {author} {\bibfnamefont {M.~R.}\ \bibnamefont
		{Evans}},\ }\href
{https://www.crcpress.com/Soft-and-Fragile-Matter-Nonequilibrium-Dynamics-Metastability-and-Flow/Cates-Evans/p/book/9781420033519}
{\emph {\bibinfo {title} {Soft and Fragile Matter: Nonequilibrium Dynamics,
			Metastability and Flow}}}\ (\bibinfo  {publisher} {CRC Press},\ \bibinfo
{year} {2000})\BibitemShut {NoStop}%
\bibitem [{SM()}]{SM}%
\BibitemOpen
\href@noop {} {}\bibinfo {note} {See Supplemental Material, which includes
	Refs.~\cite{
		Schrodinger,Smoluchowski,Redner,SD,Ptaszynski,BN,Feller,LLL,Siegert},
	for~further details}\BibitemShut {NoStop}%
\bibitem [{\citenamefont {Feller}(1957)}]{Feller}%
\BibitemOpen
\bibfield  {author} {\bibinfo {author} {\bibfnamefont {W.}~\bibnamefont
		{Feller}},\ }\href@noop {} {\emph {\bibinfo {title} {{An Introduction to
				Probability Theory and Its Applications}}}}\ (\bibinfo  {publisher} {John
	Wiley},\ \bibinfo {address} {New York},\ \bibinfo {year} {1957})\BibitemShut
{NoStop}%
\bibitem [{\citenamefont {Levitov}\ \emph {et~al.}(1996)\citenamefont
	{Levitov}, \citenamefont {Lee},\ and\ \citenamefont {Lesovik}}]{LLL}%
\BibitemOpen
\bibfield  {author} {\bibinfo {author} {\bibfnamefont {L.~S.}\ \bibnamefont
		{Levitov}}, \bibinfo {author} {\bibfnamefont {H.}~\bibnamefont {Lee}}, \ and\
	\bibinfo {author} {\bibfnamefont {G.~B.}\ \bibnamefont {Lesovik}},\ }\href
{https://doi.org/10.1063/1.531672} {\bibfield  {journal} {\bibinfo  {journal}
		{J. Math. Phys.}\ }\textbf {\bibinfo {volume} {37}},\ \bibinfo {pages} {4845}
	(\bibinfo {year} {1996})}\BibitemShut {NoStop}%
\bibitem [{\citenamefont {Bagrets}\ and\ \citenamefont {Nazarov}(2003)}]{BN}%
\BibitemOpen
\bibfield  {author} {\bibinfo {author} {\bibfnamefont {D.~A.}\ \bibnamefont
		{Bagrets}}\ and\ \bibinfo {author} {\bibfnamefont {Y.~V.}\ \bibnamefont
		{Nazarov}},\ }\href {\doibase 10.1103/PhysRevB.67.085316} {\bibfield
	{journal} {\bibinfo  {journal} {Phys. Rev. B}\ }\textbf {\bibinfo {volume}
		{67}},\ \bibinfo {pages} {085316} (\bibinfo {year} {2003})}\BibitemShut
{NoStop}%
\bibitem [{Note2()}]{Note2}%
\BibitemOpen
\bibinfo {note} {This time scale is larger than the electron-phonon
	relaxation time $\sim 10 ^{-6}$~s, and electron-electron relaxation time
	$\sim 10 ^{-9}$~s.}\BibitemShut {Stop}%
\bibitem [{\citenamefont {Siegert}(1951)}]{Siegert}%
\BibitemOpen
\bibfield  {author} {\bibinfo {author} {\bibfnamefont {A.~J.~F.}\
		\bibnamefont {Siegert}},\ }\href {\doibase 10.1103/PhysRev.81.617} {\bibfield
	{journal} {\bibinfo  {journal} {Phys. Rev.}\ }\textbf {\bibinfo {volume}
		{81}},\ \bibinfo {pages} {617} (\bibinfo {year} {1951})}\BibitemShut
{NoStop}%
\end{thebibliography}
\end{document}